\documentclass[useAMS,usenatbib]{mnras}
\usepackage[T1]{fontenc}
\usepackage{graphicx}   % Including figure files
\usepackage{amssymb}    % Extra maths symbols
\usepackage{mathtools}
\usepackage{bm} % bold italic vectors in math mode

\usepackage{subcaption}
\title[Orthogonal pulsars as a key test for pulsar evolution]
{Orthogonal pulsars as a key test for pulsar evolution}

\author[E.~M.~Novoselov, V.~S.~Beskin, A.~K.~Galishnikova, M.~M.~Rashkovetskyi  and A.~V.~Biryukov]{\parbox{\textwidth}
{E.~M.~Novoselov$^{1}$,
V.~S.~Beskin$^{2, 3}$\thanks{Contact e-mail: \href{mailto:beskin@lpi.ru}{beskin@lpi.ru}}, 
A.~K.~Galishnikova$^{2, 4}$, M.~M.~Rashkovetskyi$^{5}$ and A.~V.~Biryukov$^{6, 7}$} 
\vspace{0.4cm}\\
\parbox{\textwidth}{
$^{1}$Columbia University, Department of Physics, New York City, 116th str.  Broadway, New York, 10027, United States \\
$^{2}$Moscow Institute of Physics and Technology, Dolgoprudny, Institutsky per. 9, Moscow Region, 141700, Russia\\
$^{3}$P.N.Lebedev Physical Institute, Leninsky prosp. 53, Moscow, 119991, Russia \\
$^{4}$Department of Astrophysical Sciences, Peyton Hall, Princeton University, Princeton, NJ 08544, USA \\
$^{5}$The Raymond and Beverly Sackler School of Physics and Astronomy, Tel Aviv University, Tel Aviv, 69978, Israel \\
$^{6}$Sternberg Astronomy Institute,  Moscow State University, Universitetsky prosp. 13, Moscow, 119234,  Russia \\
$^{7}$Kazan Federal University, Institute of Physics, Kremlyovskaya str. 18, Kazan, 420008, Russia}
}

\begin{document}

\date{Accepted, Received}

%\pagerange{\pageref{firstpage}--\pageref{lastpage}} \pubyear{2008}

\maketitle

\label{firstpage}

\begin{abstract}
At present, there is no direct information about evolution of inclination angle 
$\chi$ between magnetic and rotational axes in radio pulsars. As to theoretical 
models of pulsar evolution, they predict both the alignment, i.e. evolution of 
inclination angle $\chi$ to $0^{\circ}$, and its counter-alignment, i.e. evolution 
to $90^{\circ}$. In this paper, we demonstrate that the statistics of interpulse 
pulsars can give us the key test to solve the alignment/counter-alignment problem 
as the number of orthogonal interpulse pulsars ($\chi \approx 90^{\circ}$) drastically 
depends on the evolution trajectory.  
\end{abstract}


\section{Introduction}

Fifty years of intensive researching of radio pulsars have not leaded to complete understanding of the nature of many key processes in the pulsar magnetosphere~\citep{ManTay1977, LyneG-S1998, LorKra2004}. We still know neither the mechanism of coherent radio emission nor the real structure of electric currents that are responsible for braking of neutron stars. In this article, we try to show how the existence of orthogonal radio pulsars itself may help us to clarify several important questions faced by the theory.

Remind that one of the most important unsolved questions is the problem of the evolution of the inclination angle $\chi$ between magnetic and rotational axes, i.e. between magnetic moment $\bm{m}$ and the angular velocity $\bm{\varOmega}$ (see, e.g.,~\citealt{Beskin2018}). Nowadays there are the theories  predicting both the alignment, i.e., evolution of inclination angle $\chi$ to $0^{\circ}$, and its counter-alignment, i.e., evolution to $90^{\circ}$. In what follows we call the first group as MHD model, as the alignment evolution is mainly based on the results of numerical simulations within ideal magneto-hydrodynamics~\citep{Tchetal2013}, and the second group as BGI model as the counter-alignment scenario was first proposed by~\citet{BGI84}. 

 Unfortunately, at present it is impossible to determine the evolution of inclination angle $\chi$ of individual pulsars directly from observations. As to different indirect methods, they give controversial results~\citep{Rankin1990, 1998MNRAS.298..625T, 2006ApJ...643..332F, 2008MNRAS.387.1755W, 2010MNRAS.402.1317Y, Pons}. In particular, it was found both directly (i.e., by the analysis of the $\chi$ distribution) and indirectly (i.e., from the analysis of the observed pulse width) that statistically the inclination angle $\chi$ decreases with period $P$ as the dynamical age $\tau_{\rm D} = P/{\dot P}$ increases. At first glance, these results definitely speak in favor of alignment mechanism. However, as was demonstrated by~\citet{BGI93}, the average inclination angle of pulsar population, $\langle\chi\rangle (\tau_{\rm D})$ can \emph{decrease} even if inclination angles of individual pulsars \emph{increase} with time due to dependence of so-called ``death line'' on the inclination angle $\chi$. Moreover, recently, by analyzing 45 years of observational data for the Crab pulsar, \citet{2013Sci...342..598L} found that the separation between the main pulse and interpulse increases at the rate of $~0.6^\circ$ per century implying similar growth of $\chi$ (see, however,~\citealt{2015MNRAS.453.3540A, 2015MNRAS.451..695Z}).

Thus, until now it was not possible to formulate a test that would allow to distinguish between these two models of evolution, as both models reproduce well enough the real distribution of pulsars on the $P$--${\dot P}$ diagram. For this reason, so far the mechanism for pulsar braking has remained unknown. Further we will show that statistics of orthogonal interpulse pulsars (with inclination angles close to $90^{\circ}$) may give us a key test to solve this problem as the number of interpulse pulsars directly depends on the inclination angle evolution.
 
 %\tracingall
 
\begin{table}
\caption{The number of axisymmetrical (SP) and orthogonal (DP) interpulse pulsars. Lower limit corresponds to the number of certainly defined pulsars, when different catalogues are in agreement with each other; the upper limit corresponds to the highest number that can be obtained (see Appendix~\ref{AppendixA}). When determining the percentage, the ATNF catalogue~\citep{Manetal2005} is used.}
\vspace{0.3cm}
\centering
\begin{tabular}{|r|c|c|c|}
\hline
  $P \, [\rm s]$      &  0.03--0.5  &   0.5--1.0  & $> \, $1.0    \\
\hline
${\cal N}_{\rm SP}$  &   $ 4 \div 10$    & $2 \div 3$      &  $0 \div 1$    \\  
 &   $(0.4 \div 1.0)\%$    &  $(0.3 \div 0.4)\%$      &     $(0.0 \div 0.1)\%$ \\  
${\cal N}_{\rm DP}$  &   $18 \div 26$    & $3 \div 5$       &  $0 \div 1$   \\  
 &   $(1.8 \div 2.6)\%$  &  $(0.4 \div 0.7)\%$     &  $(0.0 \div 0.1)\%$   \\  
\hline
\end{tabular}
\label{table0}
\end{table}

As is well-known~\citep{ManTay1977, LyneG-S1998, Malov1990}, the interpulse appears if we observe either two opposite poles (a double-pole or DP pulsar) or the same pole twice (a single-pole or SP pulsar); in the latter case, the two peaks correspond to the double intersection of the same hollow-cone directivity pattern. For the DP case, the inclination angle $\chi$ is close to $90^{\circ}$, while for a SP pulsar, this angle is close to $0^{\circ}$. Since the procedure for determining of the inclination angle (which is based on the analysis of the polarization properties of mean profiles) contains a number of uncertainties, different catalogues (see, e.g.,~\citealt{Maciesiak2011, MalNik2013}) give different number of orthogonal pulsars (see Table~\ref{table0} and Appendix~\ref{AppendixA}). But in any way we can be sure that there are about dozens of them, therefore several certain claims can be done.

The present paper is organized as follows. In Section~\ref{Sect:2} the necessary data on the properties of orthogonal pulsars are given. Further, in Section \ref{Sect3.1} we start with defining the regions within the polar cap where the secondary plasma is generated for nearly orthogonal pulsars. Assuming, as is usually done, that the intensity of the radio emission is proportional to the density of the outgoing plasma (in the BGI model, the radio luminosity is a fraction of $10^{-5}$ of the plasma energy flux) this allows us not only to clarify the shape of the directivity pattern for interpulse pulsars, but also define the parameters of the ``death-line'' and, in particular, its dependence on the magnetic field on the neutron star magnetic pole $B_{0}$ (Section \ref{Sect3.2}). Then, according to this information, we make an observational constraint on one of the key model parameters (Section \ref{Sect3.3}). In Section \ref{Sect:vis} we define the visibility function that connect the ``true'' distribution by physical parameters to the observed one for orthogonal interpulse pulsars. Then in Section \ref{Sect:popsynt} we use the kinetic equation approach to obtain the distributions theoretically. Finally, in Section \ref{Sect:test} we determine the number of expecting orthogonal interpulse pulsars for different models and show that such an analysis provides a test to distinguish the alignment and counter-alignment evolution scenario. The Section \ref{Sect:6.1} is a recap of previous results, while in \ref{Sect:6.2} we make a crucially important correction to BGI model and provide its main implications. Then in Section \ref{Sect:6.3} we conduct a Monte-Carlo simulation to consider the corrected evolution law more thoroughly and compare its results to observations. Finally, in Appendix we discuss the ATNF catalogue limitations and give possible ways that would allow us to overcome the incompleteness of this catalog.

It should be emphasized that such a consideration, when two-dimensional distribution of the density of the outflowing plasma over the surface of the polar cap in the vicinity of the death line is studied, is actually carried out for the first time. Until now, the efficiency of particle production has been discussed either in the one-dimensional case (see, e.g., \citealt{Shibata, Beloborodov, Tim2010, TimArons2013, TimHar2015}) or, more recently, far from the death line~\citep{B&S, Lockhart}. In any case, in these works the strong distortion of the radiation pattern of orthogonal pulsars was not discussed in detail. As a result, it is still often assumed that for orthogonal pulsars the directivity pattern has standard hollow cone structure (see e.g.~\citealt{SiMi}). It is clear that this question becomes the key one in the analysis of the visibility function of orthogonal pulsars.

%\tracingall

\section{Orthogonal interpulse pulsars}
\label{Sect:2}

To begin with, let us discuss two examples that show what kind of information we can gain from the very existence of orthogonal interpulse pulsars. For simplicity, we put the inclination angle $\chi = 90^{\circ}$ in this section.

\begin{figure}
\centering
\includegraphics[scale=0.45]{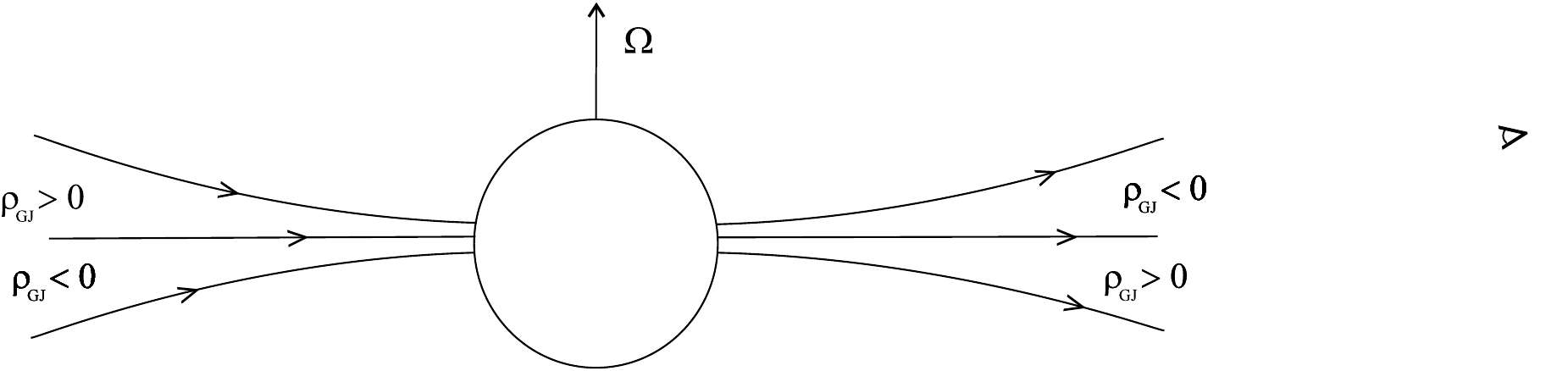}
\caption{Geometry of an orthogonal radio pulsar with inclination angle $\chi = 90^{\circ}$. Observer located over the equatorial plane detects two pulses corresponding to two areas with different signs of Goldreich-Julian charge density $\rho_{\rm GJ}$ (\ref{GJ}).}
\label{fig:1}
\end{figure}

At first, remember that morphological and polarization properties of both pulses in observable interpulse pulsars are similar to ordinary ones (see, e.g.,~\citealt{Keith2010}). Thus, the process of secondary plasma generation passes with the same pace as it is for ordinary pulsars. On the other hand, as is shown in Figure~\ref{fig:1}, main pulse and interpulse can belong to areas with different signs of~\citet{GJ1969} charge density (the one that is required to screen the longitudinal electric field)
\begin{equation}
    \rho_{GJ} = -\frac{\bm{\varOmega B}}{2\pi c},
\label{GJ}
\end{equation}
i.e., to different signs of the normal vector of the electric field in areas of plasma particles acceleration. Hence, one can conclude that the injection of particles from the star surface does not play important role in pair creation mechanism as the work function and moreover masses of electrons and protons differs significantly.

In other words, the very existence of orthogonal interpulse pulsars supports~\citet{RS1975} theory, according to which the ejection of electrons from the star surface was assumed to be inefficient, in opposition to~\citet{Arons1982} model, in which it was supposed that ejection is free. Remind that latest simulations of particle generation in polar regions of neutron stars~\citep{Tim2010, TimArons2013, TimHar2015} have actually proved that the injection from the surface does play a small role.

We know that to support the high efficiency of pair creation in the polar region of neutron star (which is necessary to detect a neutron star as a radio pulsar), the potential drop near the star surface is to be similar to one for ordinary pulsars. For Ruderman-Sutherland-type models the potential drop $\psi$ in the pair creation region can be evaluated as
\begin{equation}
\psi \approx 2 \pi \rho_{\rm GJ} H^{2},   
\label{psi}
\end{equation}
where $H$ is the inner gap height. This implies that potential drop $\psi$ (\ref{psi}) is proportional to Goldreich-Julian charge density \mbox{$\rho_{\rm GJ} \propto B \cos\theta_{b}$,} where $\theta_{b}$ is the angle between angular velocity $\bm{\varOmega}$ and magnetic field $\bm{B}$. 

On the other hand, for orthogonal interpulse pulsars $\cos\theta_{b}$ near the polar caps is rather small:
\begin{equation}
    \cos\theta_{b} \sim \left(\frac{\Omega R}{c}\right)^{1/2}.
\label{costhetab}
\end{equation}
For this reason, for interpulse pulsars their magnetic field $B_{0}$ should be much larger than for ordinary pulsars to maintain the same efficiency of pair production. As even for fast pulsars with period $P$ about $0.1-0.5$  seconds (as was shown in Table~\ref{table0}, most interpulse pulsars belong to this range) the corresponding factor $(\Omega R/c)^{-1/2} = 20$--$40$, observed interpulse pulsars must have ten times stronger magnetic field than for ordinary pulsars. Hence, to analyze their statistical properties, the magnetic field distribution should be taken into account. 

To sum up, one can conclude that detailed study of properties of orthogonal interpulse pulsars (that has not been done yet) should allow the significant advance in understanding the key processes in the neutron star magnetosphere. Moreover, as we show in what follows, a test that may answer the question about the evolution of the inclination angle can be formulated.

%\tracingall

\section{Directivity pattern of orthogonal pulsars}

\subsection{Plasma generation area}
\label{Sect3.1}

Before formulating the test itself, we need to find the shape of the directivity pattern of orthogonal pulsars, which must be associated with the region of plasma generation in their polar cap. It is an essential question, as we need to find the range of the inclination angles where both poles can be observable. The main difference with the pulsars having moderate inclination angles $\chi$ is that for almost orthogonal pulsars the particle creation is to be depressed not only near the magnetic pole where the curvature radius $R_{\rm c}$ of magnetic field lines is very large, but also near the line where the Goldreich-Julian charge density (\ref{GJ}) vanishes. Indeed, according to (\ref{psi}), in this case, the potential drop $\psi$ is too small to create pairs. Hence, the shape of the directivity pattern that highly depends on the charge density is to differ significantly from the standard hollow cone. 

To estimate the height of the inner gap $H$ (a domain with nonzero longitudinal electric field where the acceleration of primary particles occurs) we use the relations presented recently by~\citet{TimHar2015}. As was mentioned above, these expressions actually do not differ from ones obtained by~\citet{RS1975}. However we will take the dependence of charge density on the angle $\theta_b$ between magnetic field and rotational axis into consideration. It is easy to do if one change $\Omega $ to $\Omega|\cos\theta_b|$. As a result, we obtain
\begin{equation}
    H_{\rm RS} = 1.1 \times 10^4 |\cos\theta_b|^{-3/7}R_{{\rm c},7}^{2/7}P^{3/7}B_{12}^{-4/7} {\rm cm}.
    \label{HRS}
\end{equation}
Here and in the following similar expressions the magnetic field $B_{12}$ is expressed in $10^{12}$ G, pulsar period $P$ in seconds, and curvature radius $R_{{\rm c},7}$ in $10^{7}$ cm. 

Below we also use the results from the same work for multiplicity of particle generation $\lambda = n_{\rm e}/n_{\rm GJ}$:
\begin{eqnarray}
    \lambda & = & 5.4 \times 10^{4}R_{{\rm c},7}^{-3/7}P^{-1/7}B_{12}^{6/7}, \qquad B_{12} < 5, \label{L1}\\
    \lambda & = & 1.6 \times  10^{5}R_{{\rm c},7}^{-3/7}P^{-1/7}B_{12}^{-1/7}, \quad \,  B_{12} > 5.
    \label{L2}
\end{eqnarray}
As a result, we may write down the outflowing plasma number density as 
\begin{equation}
    n_{\rm e} = \lambda \frac{\Omega B |\cos\theta_b|}{2\pi c e}.
\end{equation}
Introducing now the polar coordinates ($\theta_m$, $\phi_m$) on the polar cap of the neutron star, we obtain for dipole magnetic field in the limit $\Omega R/c \ll 1$ 
\begin{equation}
    \theta_b = \chi - \frac{3}{2}\theta_m \sin\phi_m.
\end{equation}
Accordingly, for dipole magnetic field the curvature radius $R_{\rm c}$ at the star surface looks like
\begin{equation}
    R_{\rm c} = \frac{4}{3}\frac{R}{\theta_m}.
\end{equation}
Thus, using relations (4)--(9) one can find the distribution of the the number density $n_{\rm e}$ of the outflowing plasma within open field line region. 

It is important to remember that equation (\ref{psi}) was obtained on the assumption that the height of the accelerating area $H$ is less that its width. Usually, the polar cap radius $R_{\rm cap} \approx (\Omega R/c)^{1/2} R$ is used as a characteristic size. And the equality $H_{\rm RS} = R_{\rm cap}$ is taken as a condition of the "death-line" on $P-\dot{P}$ diagram~\citep{ManTay1977, LyneG-S1998}. But for orthogonal pulsars, we must clarify this criterion. Below, as a condition for the applicability of the one-dimensional approximation, we put
\begin{equation}
    H_{\rm RS}  < \mathcal{R}_{\rm min}
\end{equation}
where $\mathcal{R}_{\rm min}$ is the distance towards nearest characteristic point or lines: magnetic pole, edge of the polar cap or the line, where $\rho_{\rm GJ} = 0$. In more detail, we put $\lambda = \lambda_{\rm RS}$ \mbox{(\ref{L1})--(\ref{L2})} for $H_{\rm RS}  < \mathcal{R}_{\rm min}$, and use linear function \mbox{$\lambda = K \mathcal{R}_{\rm min}$} matching it to the boundary $H_{\rm RS} = \mathcal{R}_{\rm min}$ so that $\lambda$ vanishes at the characteristic lines.

\begin{figure}
    \begin{subfigure}{0.55\textwidth}
        \centering
        \includegraphics[width=0.9\linewidth]{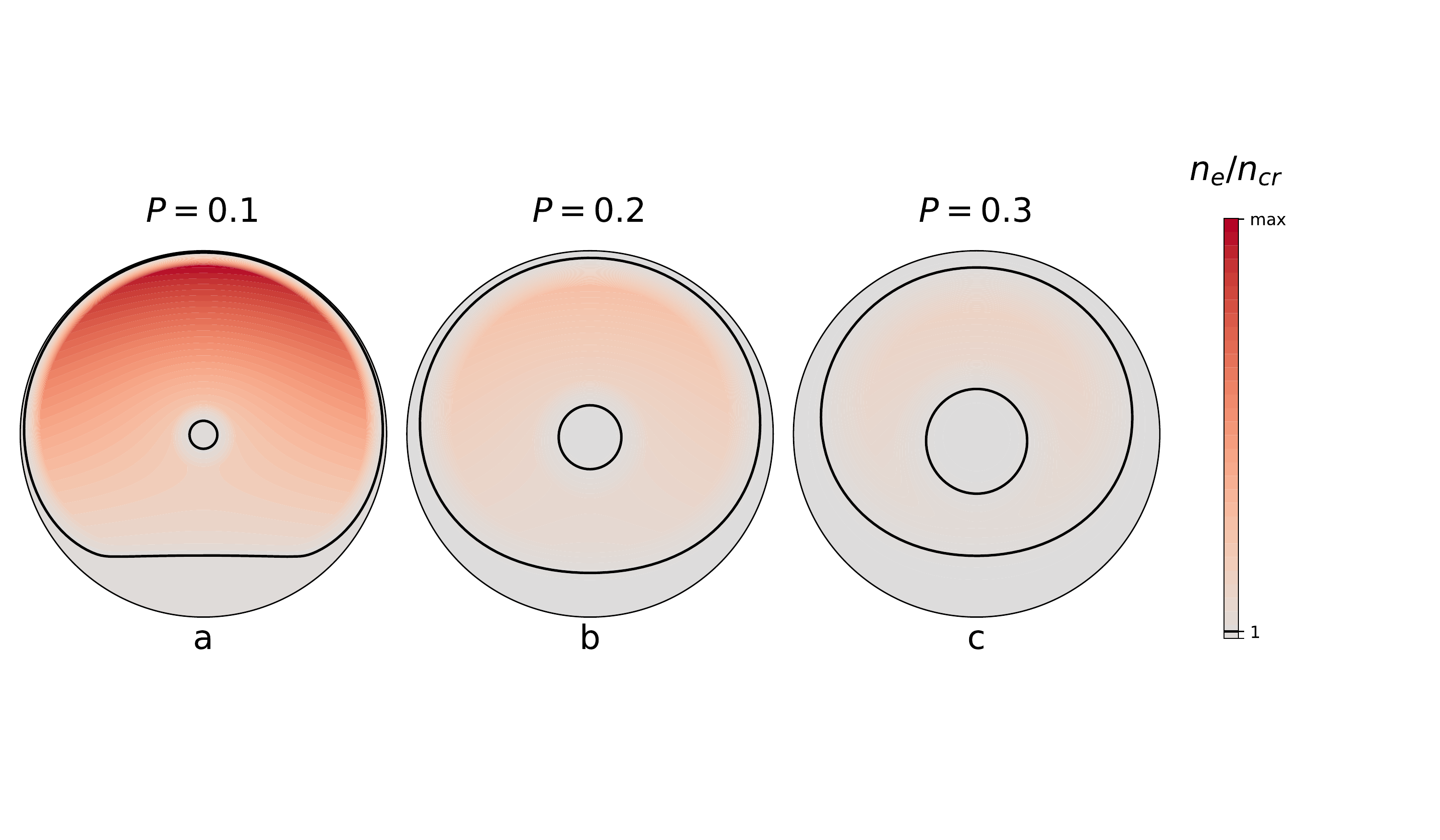}
        \caption{Different periods $P$ ($\chi = 85^{\circ}$, $B_{12} = 2$).}
        \label{fig:cap1}
    \end{subfigure}%
    \newline
    \begin{subfigure}{0.55\textwidth}
        \centering
        \includegraphics[width=0.9\linewidth]{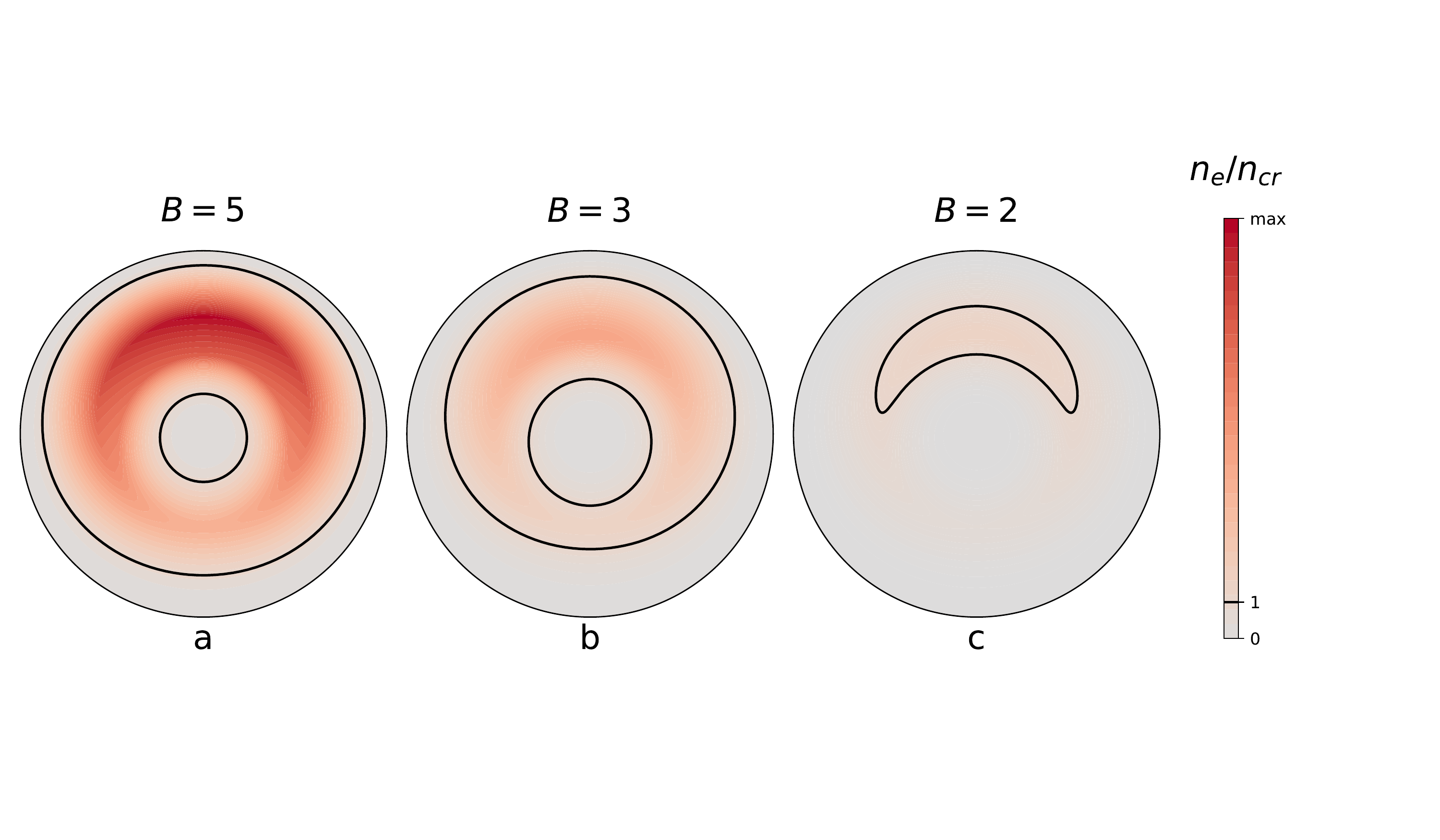}
        \caption{Different magnetic fields $B_{12}$ ($\chi = 85^{\circ}$, $P = 0.5$).}
        \label{fig:cap2}
    \end{subfigure}%
    \newline
    \begin{subfigure}{0.55\textwidth}
        \centering
        \includegraphics[width=0.9\linewidth]{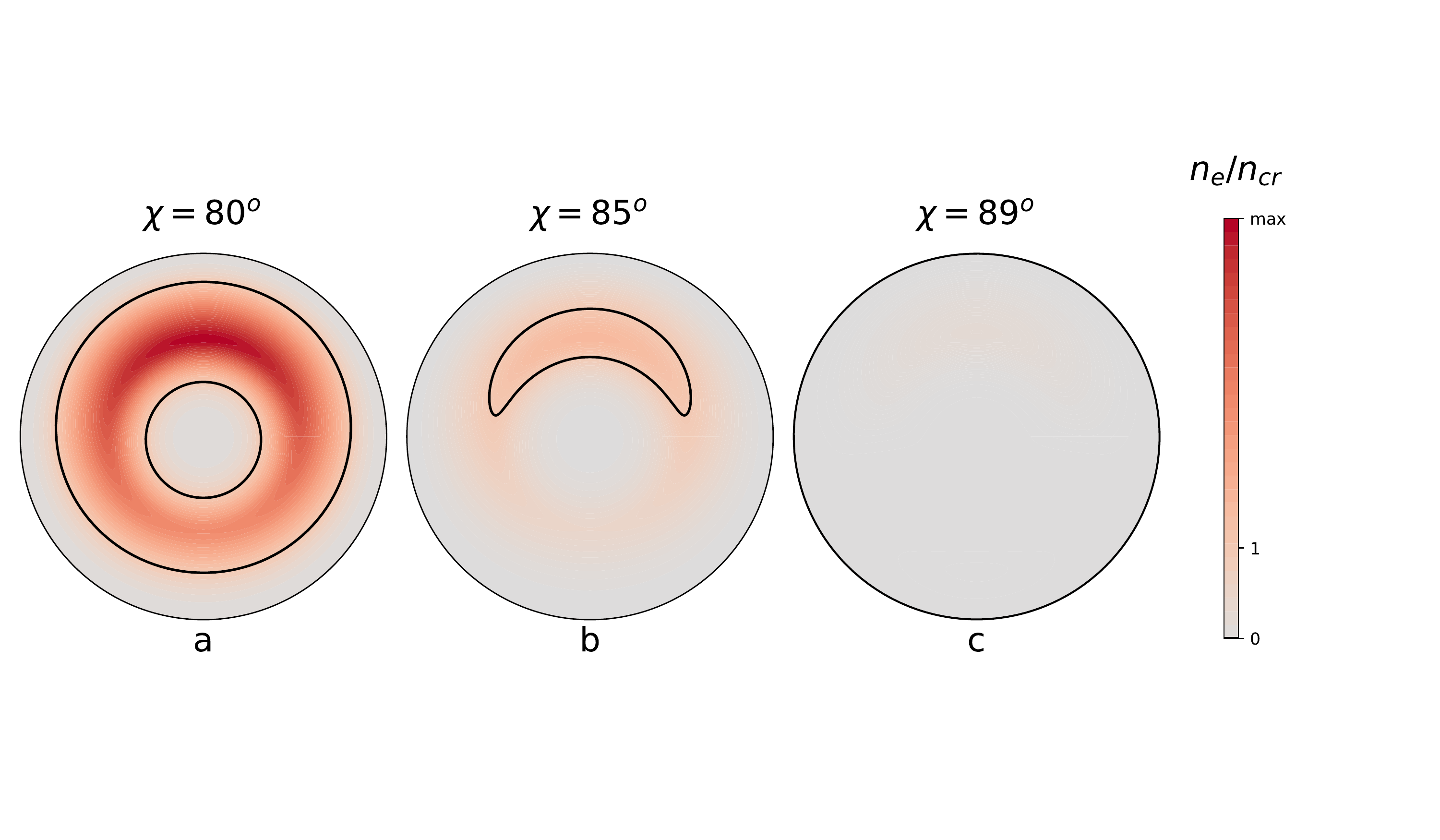}
        \caption{Different inclination angles $\chi$ ($P = 0.5$, $B_{12} = 2$).}
        \label{fig:cap3}
    \end{subfigure}%
    \caption{%\AG{Changed structure of figures, old version is commented below} 
    Geometry of plasma generation region for different parameters. Critical number densities $n_{\rm cr}$, which determine the shape of the directivity pattern, are shown in bold line.}
    \label{fig:caps}
\end{figure}

In Figure~\ref{fig:caps} we show the areas of plasma generation region for different periods $P$, magnetic fields $B_{0}$, and inclination angles $\chi$. Level lines correspond to number density $n_{\rm e}$ of outflowing plasma; critical number density $n_{\rm cr}$ determining the shape of the directivity pattern will be found in the next Subsection. Here we make an assumption that the polar cap has circle shape with its radius $R_{\rm cap} = f_{\ast}^{1/2}(\Omega R/c)^{1/2}R$, where the dimensionless polar cap area $f_{\ast}$ belongs to interval between 1.59 while $\chi = 0^{\circ}$ and 1.96 while $\chi = 90^{\circ}$~\citep{BGI83, Gralla}. As one can see, while the inclination angle $\chi$ goes to $90^{\circ}$, geometry of plasma generating areas become more and more complicated. For large periods $P$ and small magnetic fields $B_{0}$ generation of secondary plasma obviously brakes down.

\subsection{``Death line'' and directivity pattern}
\label{Sect3.2}

Now, we have to determine so-called ''death line'', i.e., the boundary of the area of pulsar parameters that admit effective secondary plasma generation. In what follows we assume that it is this condition that determines the boundary of the directivity pattern of the radio emission.

As was shown by~\citet{BGI93} (see also~\citealt{Arzamassky2017}), within Ruderman-Sutherland model, the condition for the efficient secondary plasma generation which is necessary for the generation of radio emission, 
can be written down as $Q_{\rm BGI} < 1$ , where 
\begin{equation}
    Q_{\rm BGI} = A \, P^{15/14}B_{12}^{-4/7}\cos^{2d-2}\chi.
\label{QBGI}
\end{equation}
Here $A \approx 1$ and $d \approx 0.75$. Since the accuracy of determining the value of $A$ is actually small, in what follows we assume that $A$ lies in the range from 0.5 to 2. To correlate this value with our numerical results we suppose that generation of radio emission takes place only for large enough number density of the outflowing plasma $n_{\rm e} > n_{\rm cr}$. 
 
\begin{figure}
    \begin{subfigure}{0.4\textwidth}
        \centering
        \includegraphics[width=0.9\linewidth]{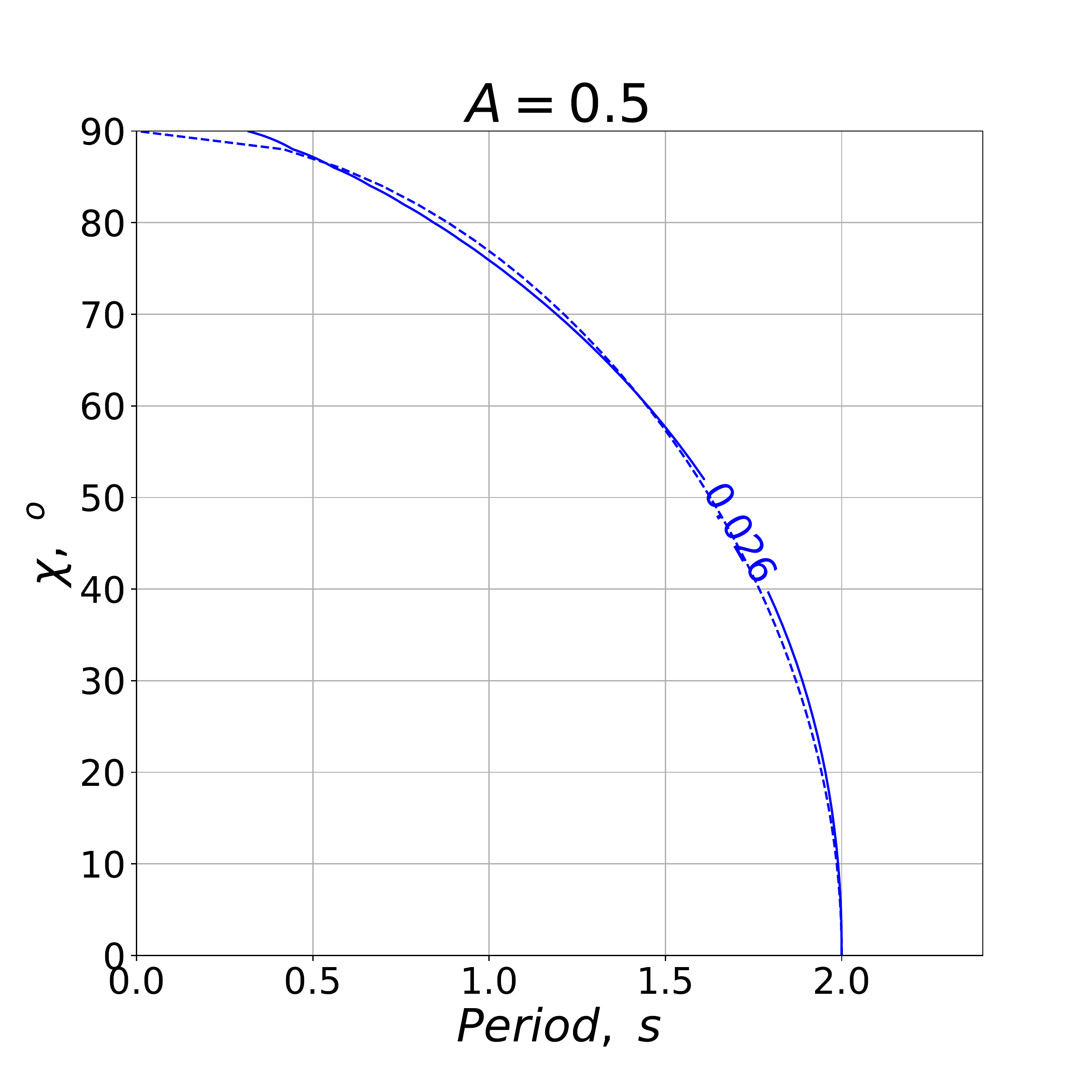}
    \end{subfigure}%
    \newline
    \begin{subfigure}{0.4\textwidth}
        \centering
        \includegraphics[width=0.9\linewidth]{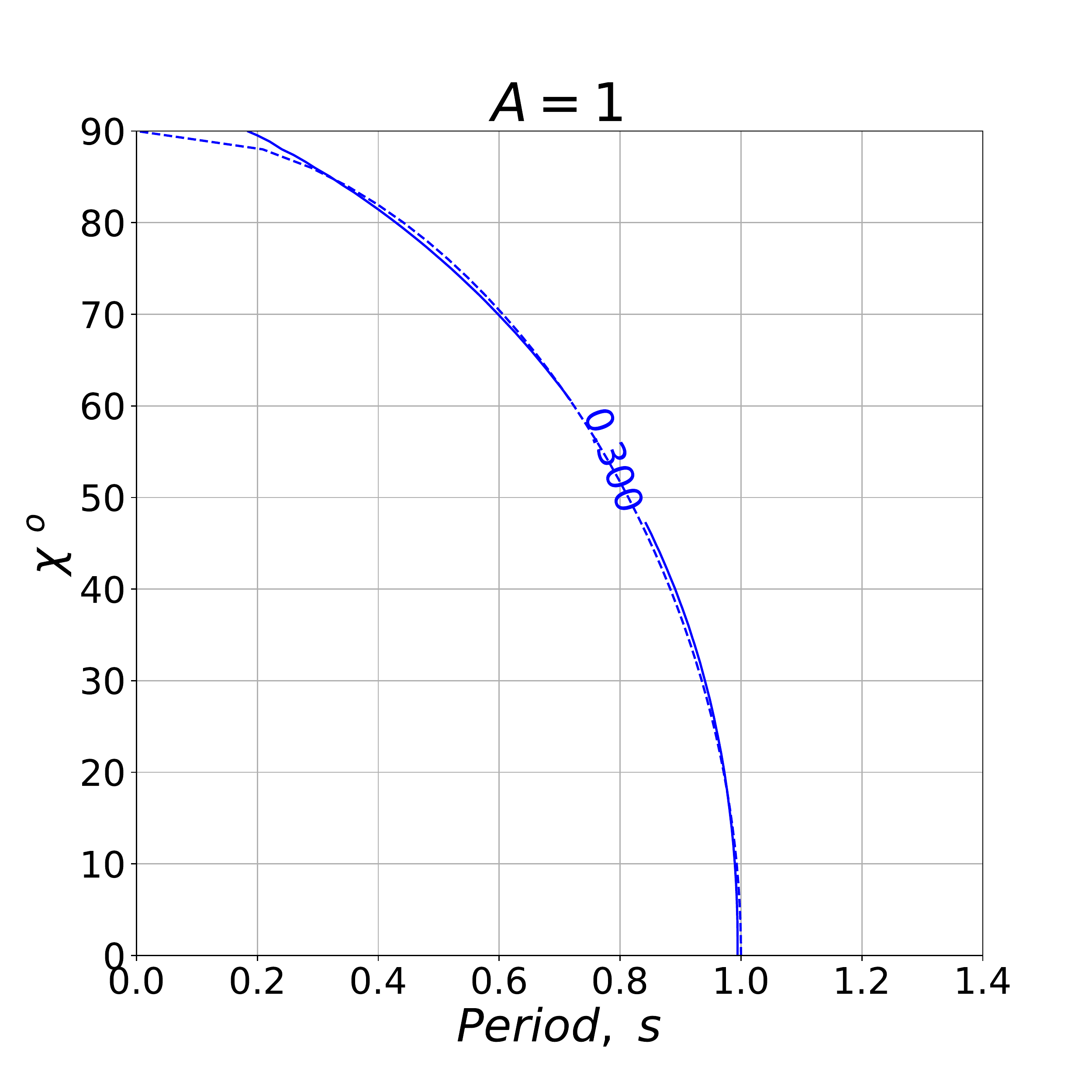}
    \end{subfigure}%
    \newline
    \begin{subfigure}{0.4\textwidth}
        \centering
        \includegraphics[width=0.9\linewidth]{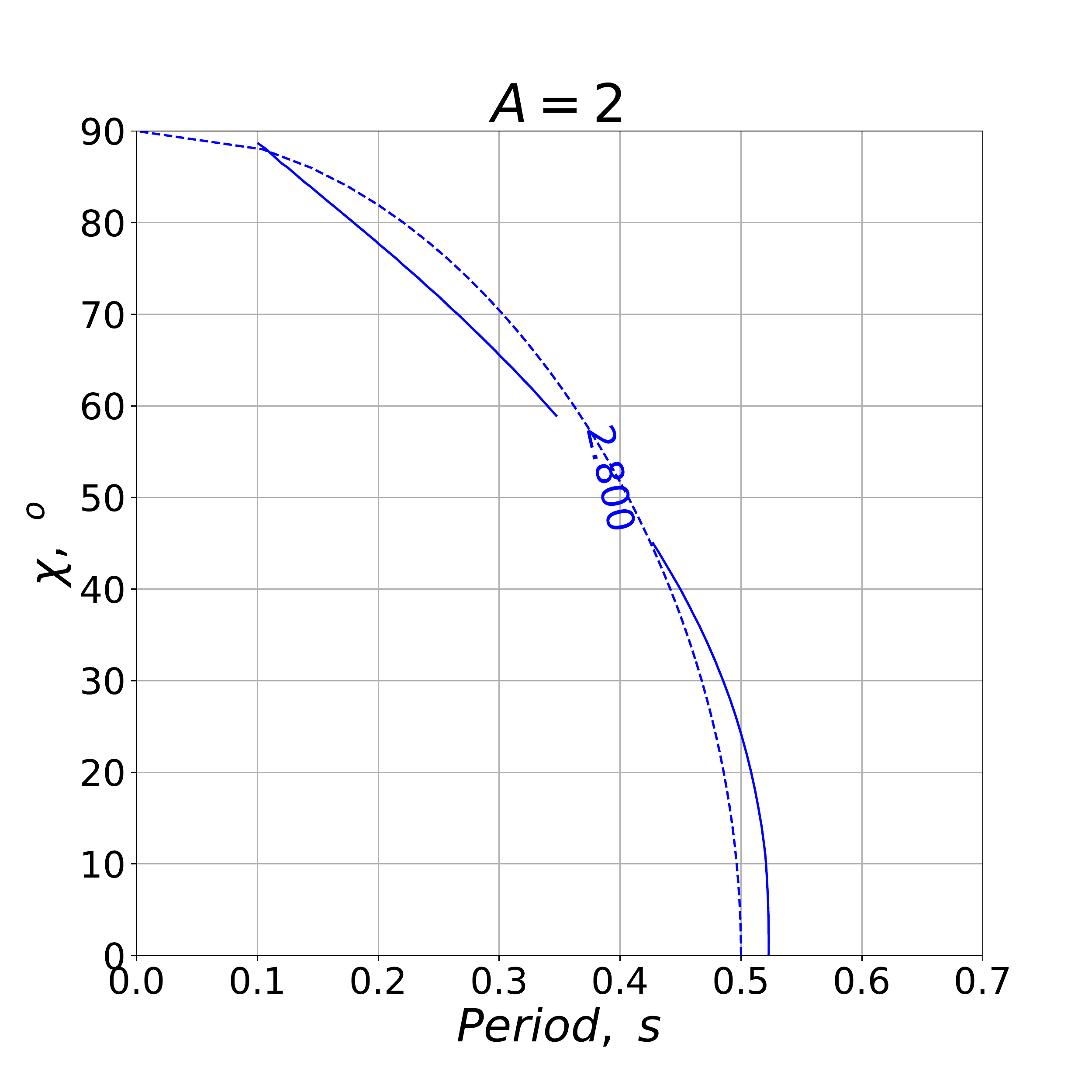}
    \end{subfigure}%
    \caption{Theoretical (\ref{QBGI}) (dashed curve) and numerical (solid curve) ``death lines" for different coefficient $A$. Appropriate values of $n_{\rm cr}$ are given in Table~\ref{table1}.}\label{fig:5}
\end{figure}

\begin{table}
\caption{Values of $n_{\rm cr}$ for three different values of $A$. }
\vspace{0.3cm}
\centering
\begin{tabular}{|c|c|c|c|}
\hline
A &  0.5  &   1.0  & 2.0      \\
\hline
$n_{\rm cr} \, (10^{15} \, {\rm cm}^{-3})$  & 0.026 & 0.3 &  2.8    \\  
  \hline
\end{tabular}
\label{table1}
\end{table}

As one can see from Fig.~\ref{fig:5}, the death line obtained numerically fits the theoretical $\theta$-dependence with an accuracy about 10\%. We take here magnetic field $B_{12} = 1$ as a clear mean value. Appropriate values $n_{\rm cr}$ are given in Table~\ref{table1}. Thus, we can use these critical number densities to determine the transverse structure of effective pair generation region for orthogonal pulsars (and, hence, the shape of their directivity pattern). In Figure~\ref{fig:caps} the corresponding critical densities are shown in bold line.
%Mean magnetic field for two models: MHD and BGI - are 0.6 and 1.4 correspondingly, so their mean the value is 1.

\subsection{Observational constraint for the value $A$}
\label{Sect3.3}

It is important that the value of $A$ can be constrained from observations while considering a subset of pulsars with measured periods $P$, their derivatives $\dot P$ and inclination angles $\chi$. Namely, pulsar death line equation $Q_{\rm BGI} = 1$ (\ref{QBGI}) can be rewritten in the form  
\begin{equation}
    \cos^{7/15}\chi = P \cdot ( A^{14/15} B_{12}^{-8/15} ),
    \label{eq:death_line_angle}
\end{equation}
where $B_{12}$ is model-consistent estimate of pulsar magnetic field taken again in $10^{12}$ G. Therefore, the ``death line'' of pulsars is conveniently depicted on the plane of the inclination angles and  the modified periods $P \cdot (A^{14/15} B_{12}^{-8/15})$.  The value of magnetic field $B_{12}$ explicitly depends on $P$, $\dot P$, and $\chi$ and can be directly obtained from pulsar spin-down law.

\begin{figure*}
    \begin{subfigure}{0.45\textwidth}
        \centering
        \includegraphics[width=2.2\linewidth]{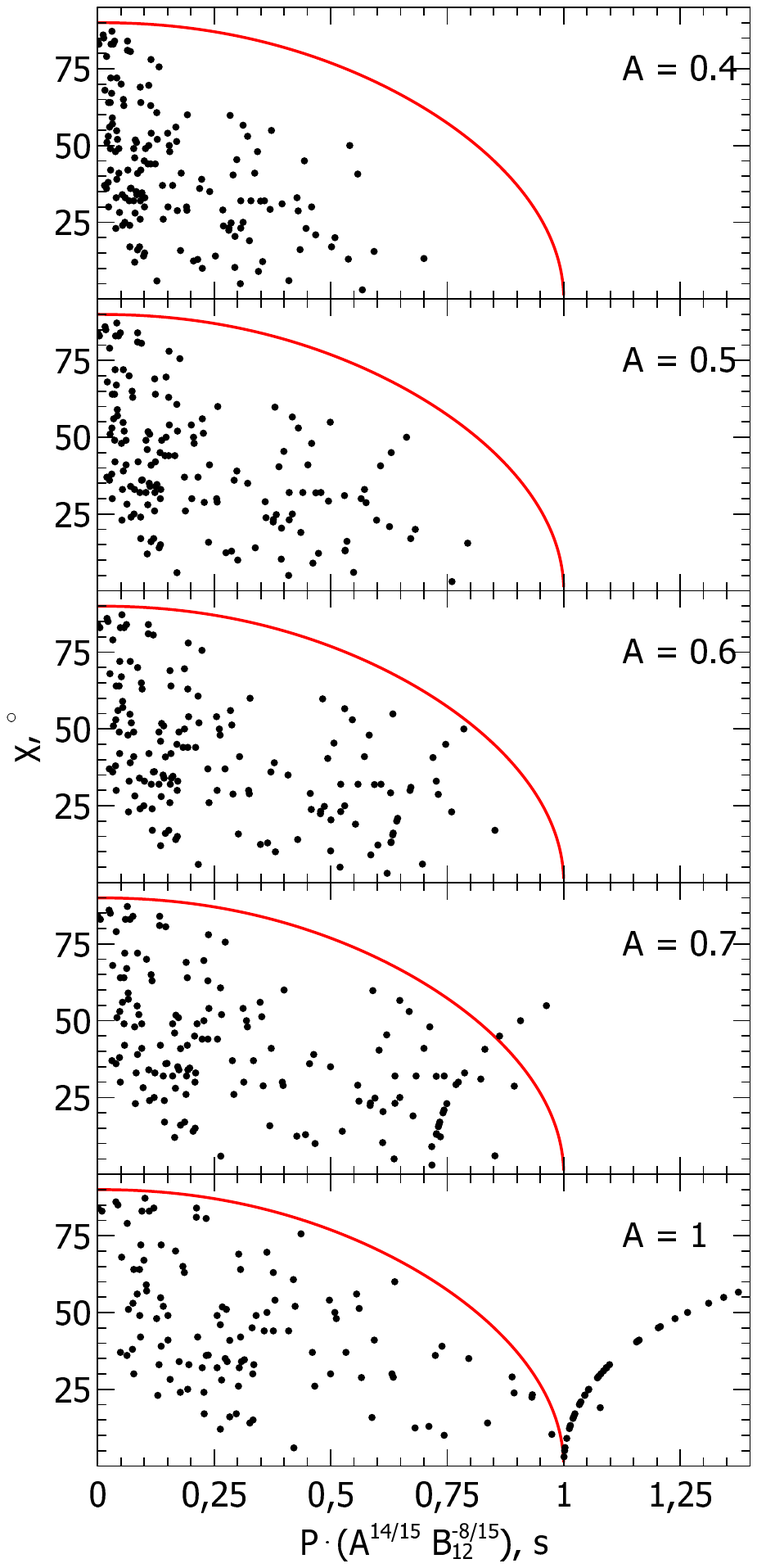}
        \caption{BGI model.}
        \label{fig:BGI_death_line_test}
    \end{subfigure}%
    \quad
    \begin{subfigure}{0.45\textwidth}
        \centering
        \includegraphics[width=2.2\linewidth]{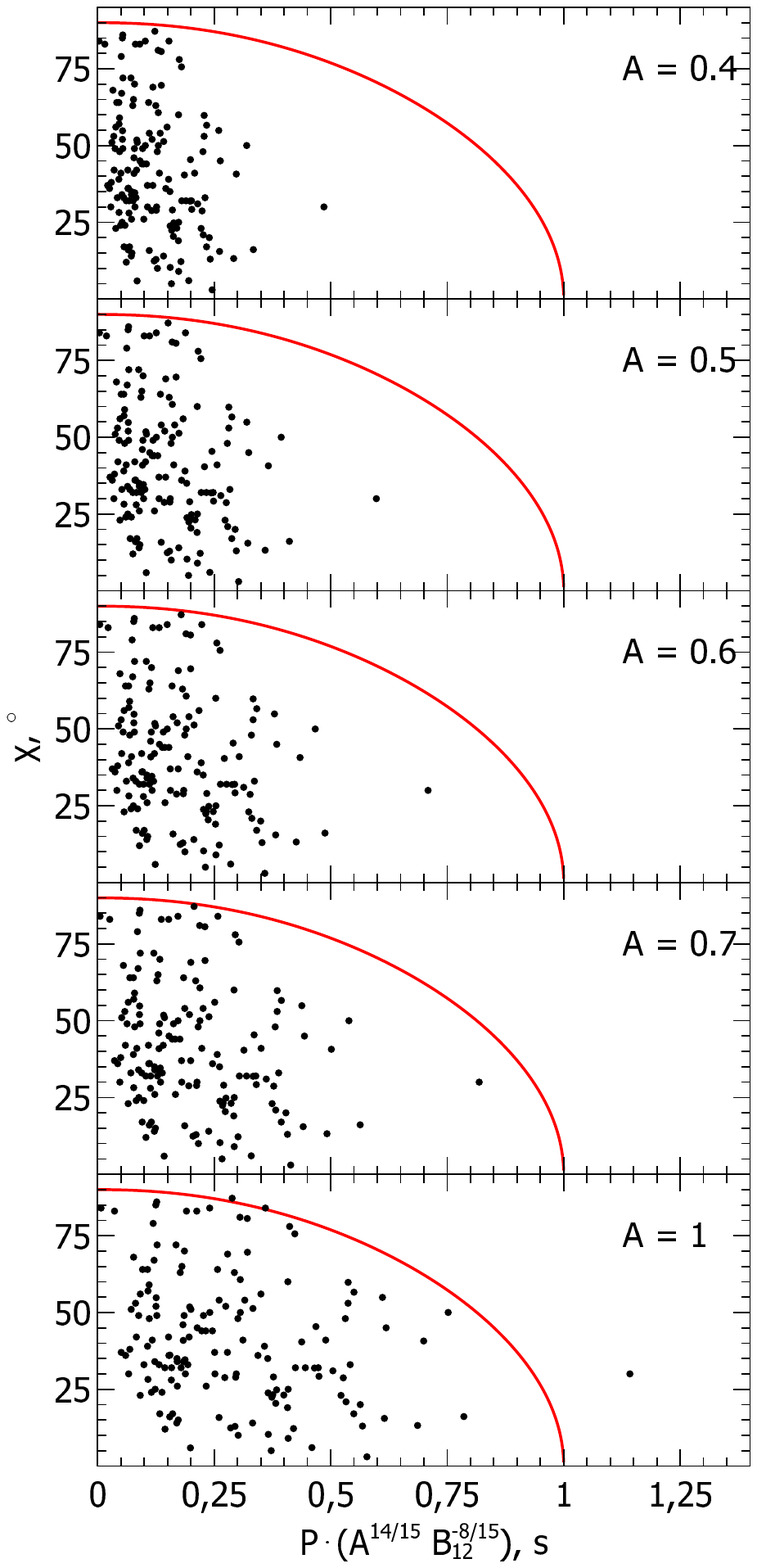}
        \caption{MHD model.}
        \label{fig:MHD_death_line_test}
    \end{subfigure}%
    \caption{Observational test for the death line equation (\ref{eq:death_line_angle}) using 153 pulsars with known inclination angles $\chi$ for different models. The ``death line'' is shown by red solid line, while pulsar positions within these plots depend on the adopted value of $A$. As one can see, the values $A \approx 0.6-0.7$ seem optimal for matching the border of pulsar cloud for both models.}
    \label{fig:death_line_test}
\end{figure*}

We considered 153 normal pulsars with inclination angles $\chi$ evaluated by \citet{LyneManchester1988_Shapes} and \citet{Rankin1993_ConalEmission} which magnetic fields were estimated within both BGI and MHD approaches (see Appendix~\ref{AppendixB}). We adopted five reasonable values for $A = 0.4, 0.5, 0.6, 0.7$ and $1.0$ and tested whether these pulsars satisfy the death line condition (\ref{eq:death_line_angle}). As one can see from Figure~\ref{fig:death_line_test}, for both models the values $A \lesssim 0.5$ presume a remarkable gap between the cloud of pulsars and proposed ``death line''. On the other hand, the values $A > 0.7$ presume significant number of objects beyond the ``death line''. 

Of course, the data used above cannot be regarded as accurate due to the measurement errors in pulsar inclination angles. And the problem here is that most of observers ignore measurement errors for $\chi$ since that are affected by significant systematic uncertainties. Indeed, estimate of $\chi$ strongly depend on the adopted model of pulsar emission geometry. Nevertheless, we believe, that estimations made by \cite{LyneManchester1988_Shapes} and  \cite{Rankin1993_ConalEmission} are relevant and their systematic errors, being unknown, still significantly less, than the whole scatter of $\chi$ throughout the pulsar subset. Thus, we conclude that $A$ can be constrained as $0.6$--$0.7$ for real pulsars. 

\section{Visibility function of the interpulse pulsars}
\label{Sect:vis}

In this section we determine the beam visibility function of interpulse pulsars $V_{\rm beam}^{\rm vis}$, that obviously plays the main role in their statistics. For ordinary pulsars it is determined by the width of the directivity pattern $W_{\rm r}$, i.e., by the radiation radius $r_{\rm rad}$, as for dipole magnetic field~\citep{ManTay1977, LyneG-S1998} 
\begin{equation}
    W_{50} = \frac{3}{2} \, \left(f_{\ast} \frac{\Omega R}{c}\right)^{1/2} \left(\frac{r_{\rm rad}}{R}\right)^{1/2}.
\label{W50}
\end{equation}
Here and below $W_{50}$ corresponds to the pulse width at the 50\% intensity level usually presented in catalogs (factor $3/2$ is well-known broadening of the dipole magnetic field). Accordingly, we suppose that the total width $W_{\rm r} = 2 W_{50}$.

Remember that observable width of the mean pulse depends on the inclination angle $\chi$
\begin{equation}
    W_{50}^{\rm obs} = \frac{W_{50}}{\sin\chi},
\label{W50_obs}
\end{equation}
where, according to (\ref{W50}), $W_{50}$ can be present as
\begin{equation}
    W_{50} = \frac{W_{0}}{\sqrt{P}}.
\label{W_50}
\end{equation}
In what follows $W_{50}$ is taken in degrees as it will be more suitable in further calculations.

\begin{figure}
\centering
\includegraphics[scale=0.1]{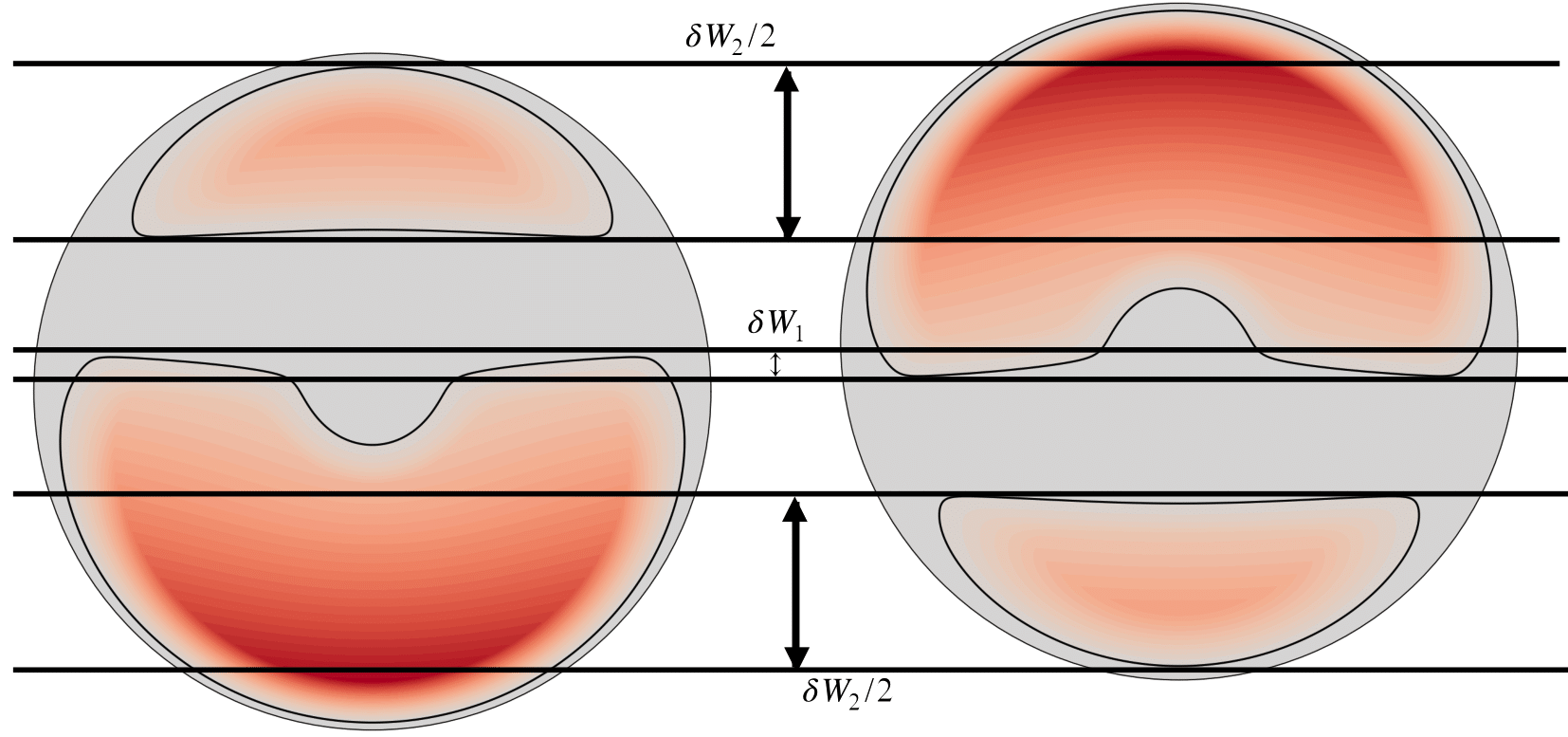}
\caption{Two cases when it is possible to observe interpulse for orthogonal pulsars.}
\label{fig:6}
\end{figure}
 
For ordinary pulsars (i.e., for pulsars with inclination angles $\chi < 90^{\circ}$) the visibility function is
\begin{equation}
V_{\rm beam}^{\rm vis} = \sin\chi \, W_{\rm r}.  
\label{Vbeam}
\end{equation}
For orthogonal interpulse pulsars we have to replace $W_{\rm r}$ with the width of the area $\delta W$ where both two poles can be observable: $V_{\rm beam,90}^{\rm vis} = \delta W$. In Fig.~\ref{fig:6} two possible realizations are shown. In the first case, when the inclination angle $\chi$ is not so close to $90^{\circ}$, it is possible to observe interpulse that correspond to the same sign of Goldreich-Julian charge density $\rho_{\rm GJ}$ (\ref{GJ}). On the contrary, when $\chi \approx 90^{\circ}$ we have to observe the regions that correspond to different signs of $\rho_{\rm GJ}$. Both cases can be implemented under certain conditions simultaneously. The total visibility function can be determined as $\delta W = \delta W_1 + \delta W_2$.

Certainly, the directivity pattern (and, therefore, the visibility function $\delta W$) depends on the generation level $r_{\rm rad}$. Below we consider two cases: $r_{\rm rad} = 5 \, R$ and $r_{\rm rad} = 7 \, R$. According to (\ref{W50}) we have \mbox{$W_{0} = 3^{\circ}$} for $r_{\rm rad} = 5 \,R $,  and $W_{0} = 5^{\circ}$ for $r_{\rm rad} = 7 \, R$ correspondingly (sf.~\citealt{Rankin1990, Maciesiak2011}). Assuming now again that the generation region of radio emission repeats the shape of the plasma generation domain, we can reconstruct the directivity pattern by transferring corresponding plasma generation profiles shown in Fig.~\ref{fig:caps} along dipole magnetic field from the neutron star surface $r = R$ to the generation level $r = r_{\rm rad}$.

\begin{figure}
    \begin{subfigure}{0.45\textwidth}
        \minipage{0.5\textwidth}
            \includegraphics[width=\linewidth]{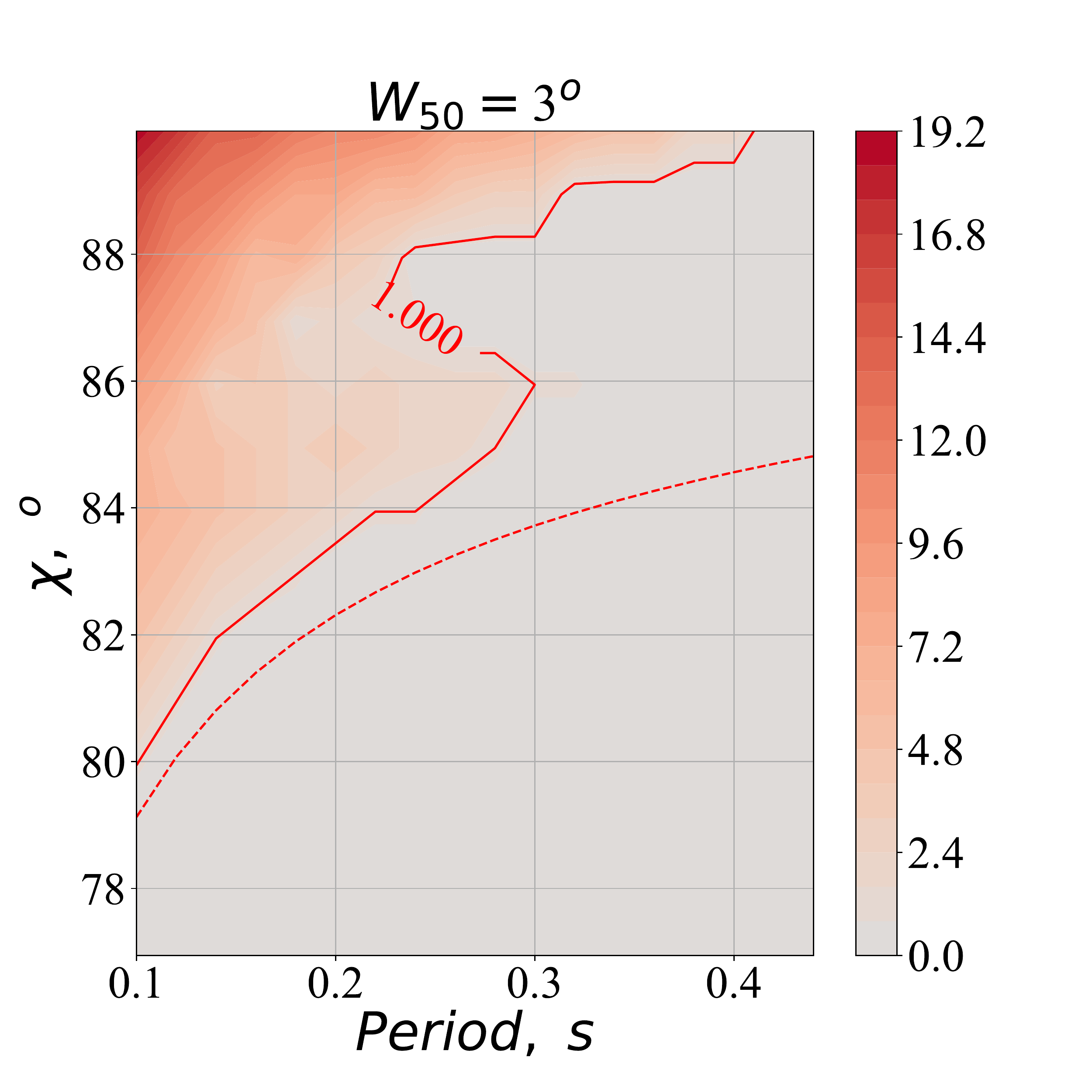}
            \endminipage\hfill
            \minipage{0.5\textwidth}
            \includegraphics[width=\linewidth]{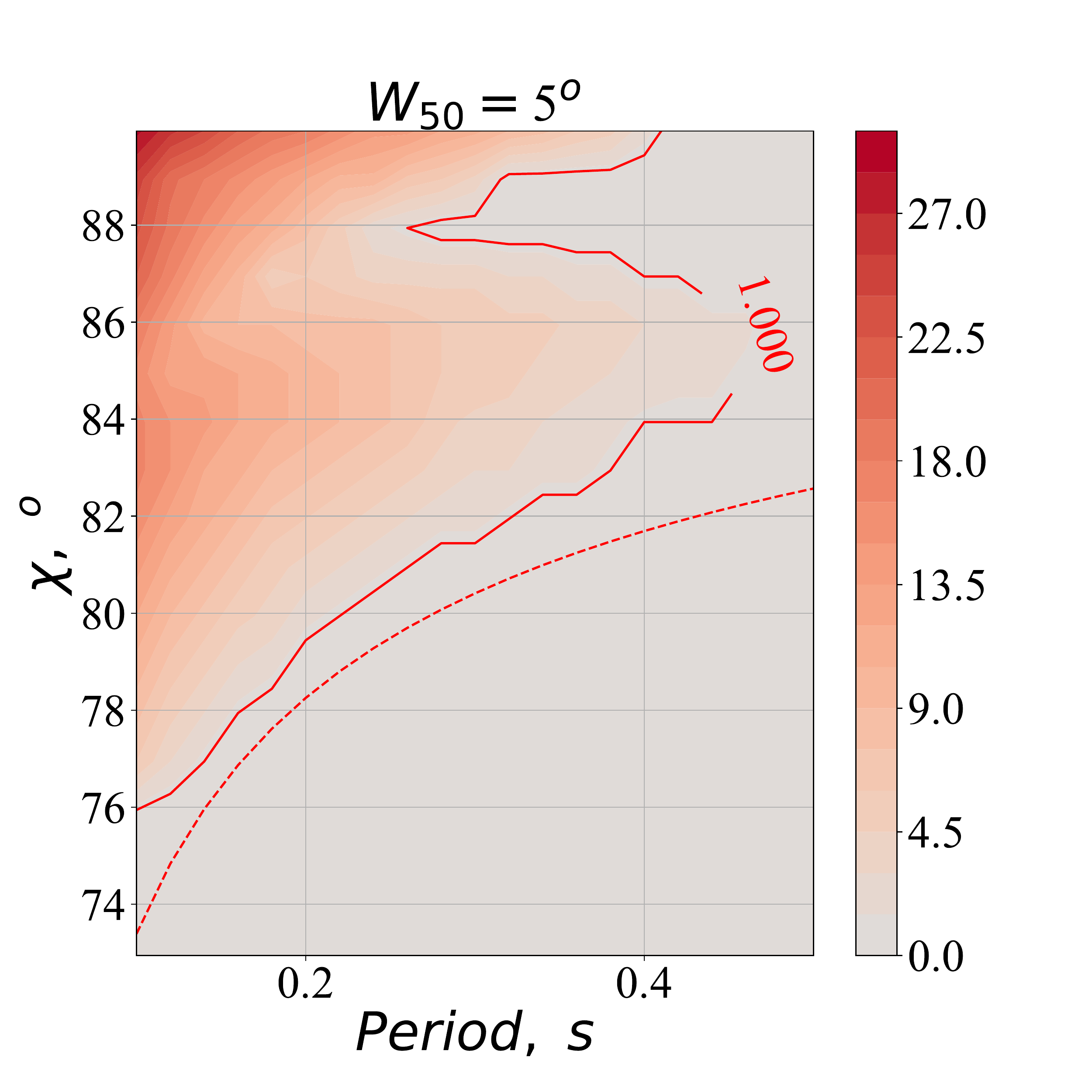}
            \endminipage\hfill
        \caption{$A = 0.5$.}
        \label{fig:7}
    \end{subfigure}%
    \newline
    \begin{subfigure}{0.45\textwidth}
        \minipage{0.5\textwidth}
            \includegraphics[width=\linewidth]{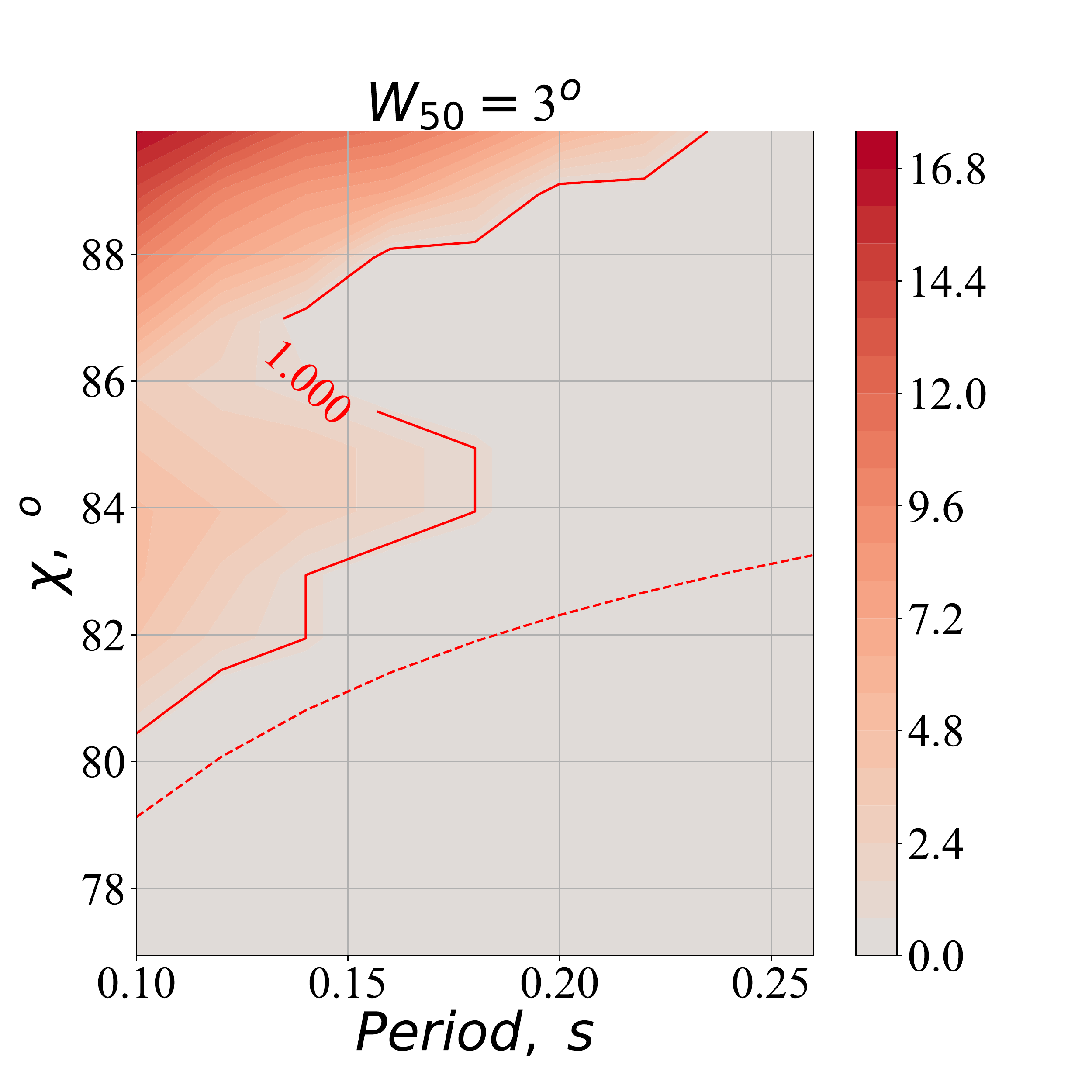}
            \endminipage\hfill
            \minipage{0.5\textwidth}
            \includegraphics[width=\linewidth]{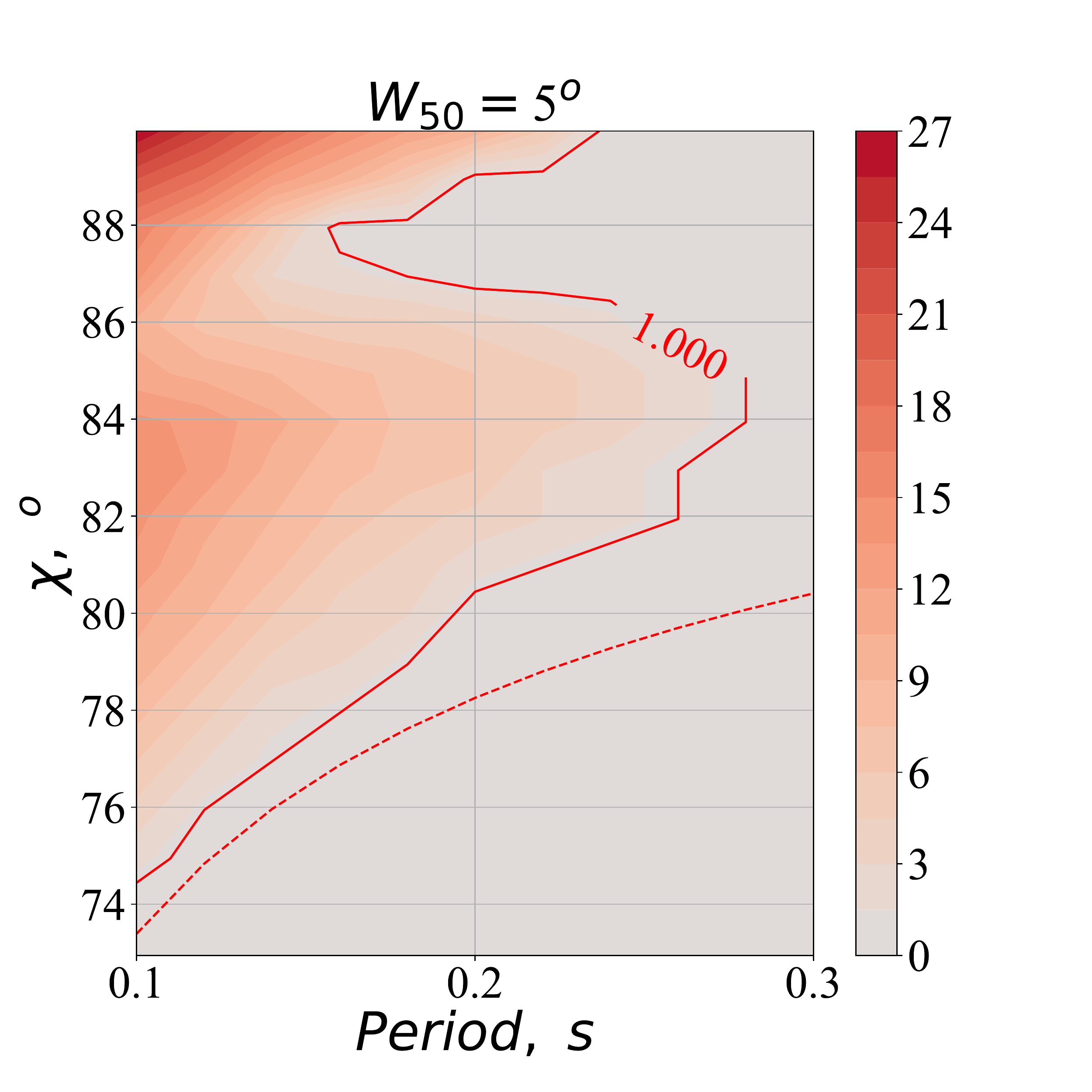}
            \endminipage\hfill
        \caption{$A = 1$.}
        \label{fig:8}
    \end{subfigure}%
    \newline
    \begin{subfigure}{0.45\textwidth}
        \minipage{0.5\textwidth}
            \includegraphics[width=\linewidth]{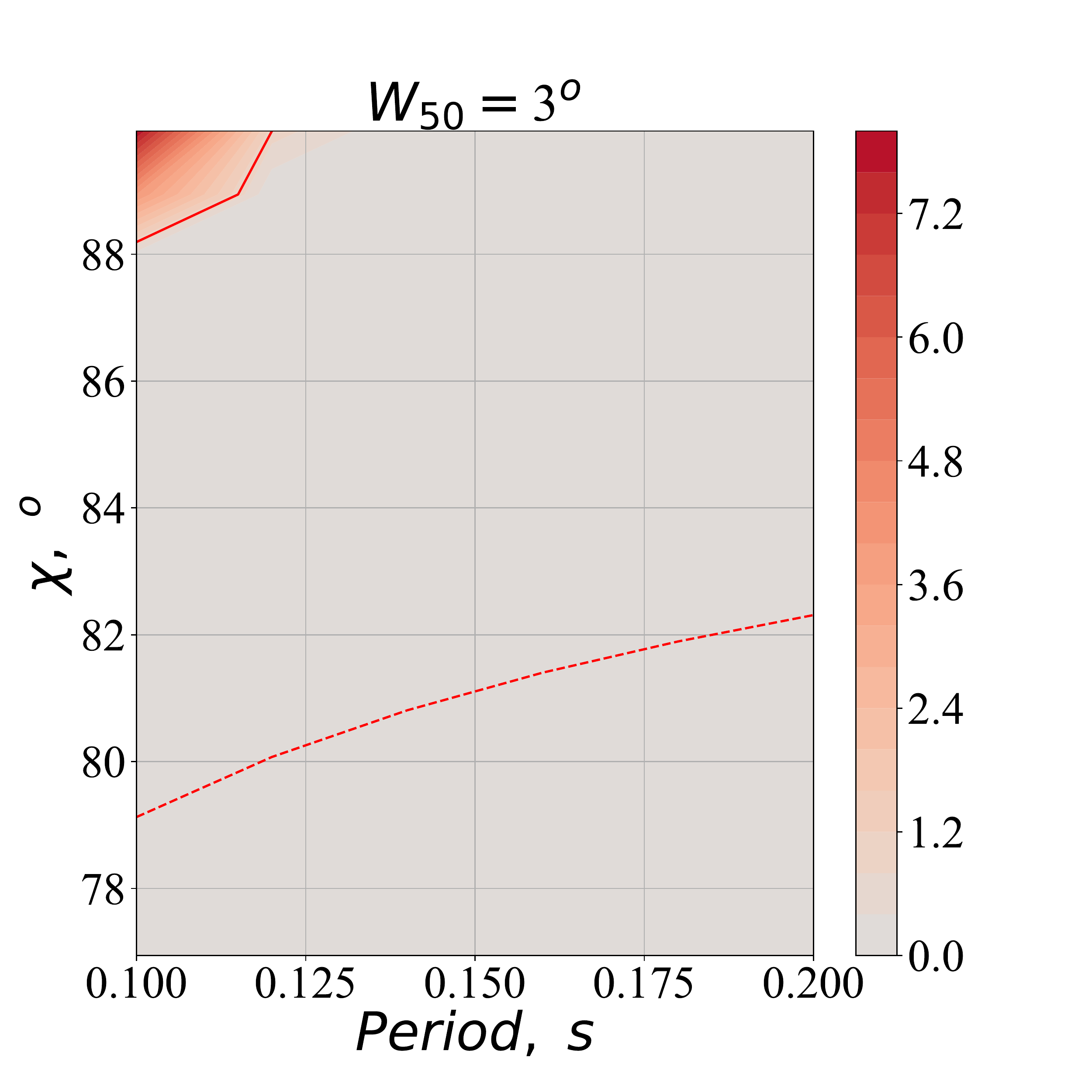}
            \endminipage\hfill
            \minipage{0.5\textwidth}
            \includegraphics[width=\linewidth]{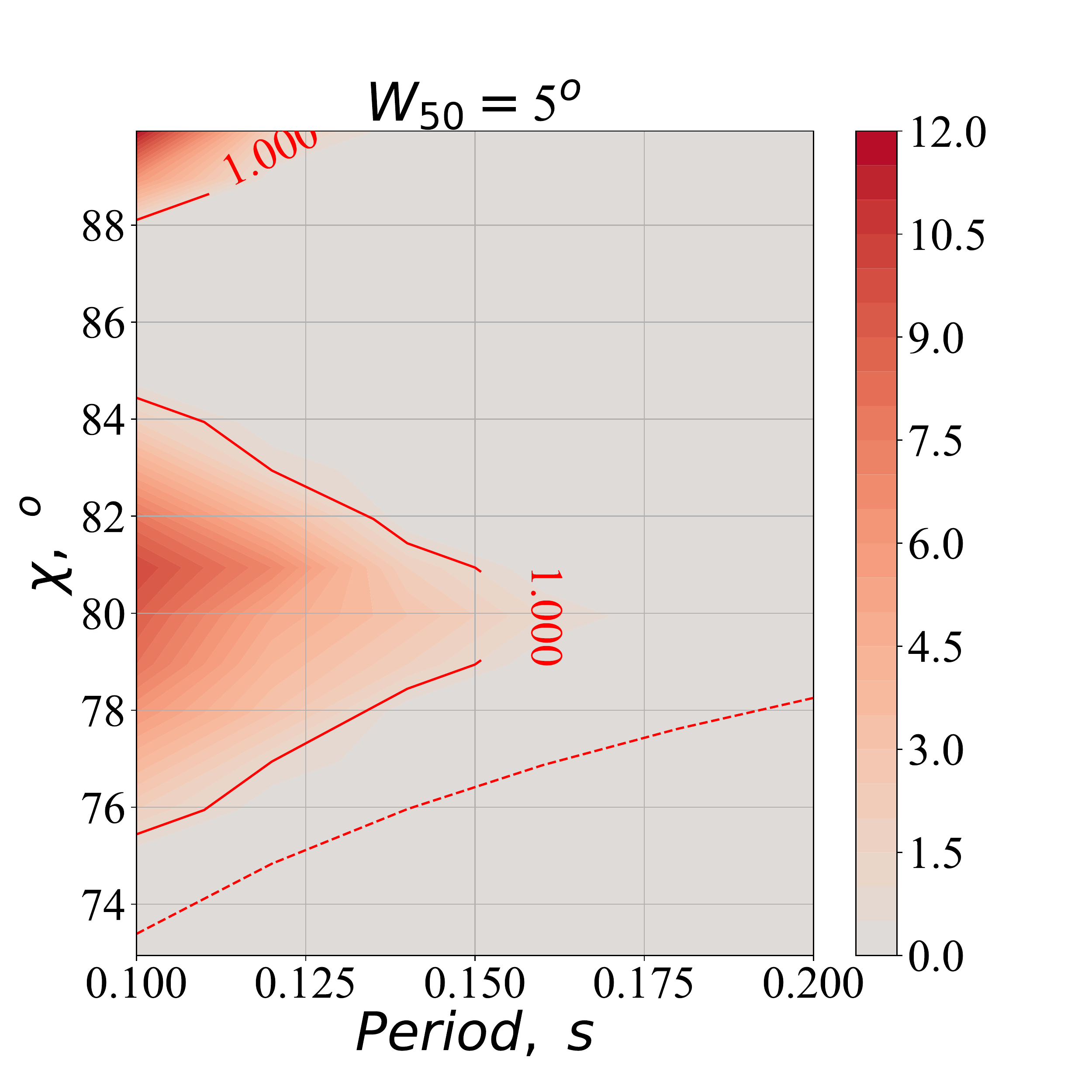}
            \endminipage\hfill
        \caption{$A = 2$.}
        \label{fig:9}
    \end{subfigure}%
    \caption{Interpulse visibility function $\delta W$ for two different window width $W_{0}$ (corresponding to two different generation level $r_{\rm rad}$) and different $A$ for magnetic field $B_{12} = 1.5$. Dotted line corresponds to the width of the total directivity pattern $2\,W_{50}$.}\label{fig:geom}
\end{figure}

In Fig.~\ref{fig:7}--\ref{fig:9} we present the visibility functions $\delta W$ for different pulsar parameters. As one can see, for ordinary magnetic field $B_{0} \sim 10^{12}$ G the possibility to observe the interpulse is limited to very small periods $P \leq 0.2$ s only. Observation of the interpulse for pulsars with periods of $P$ of the order of 0.5 s becomes possible, as was noted above, only for large enough magnetic fields $B_{0} \sim 10^{13}$ G, which is very rare. 

In particular, for the case when the main pulse and interpulse correspond to different signs of the charge density $\rho_{\rm GJ}$, the ``death line'' does not depend on the width of the directivity pattern. It should be so, since at $\chi = 90^{\circ}$ the possibility to observe radiation from one pole means that the second one will be also registered. In other words, the ``death line'' is associated only with the cessation of secondary plasma generation, and not with the observer’s exit from the directivity pattern. For this realization one can obtain numerically for the maximum period $P_{\rm cr}^{\rm num}$ (see also Fig.~\ref{fig:7}--\ref{fig:9})
\begin{equation}
P_{\rm cr}^{\rm num} \approx 0.15 \, A^{-0.76} \, B_{12}^{0.43}.
\label{Pcr_num}
\end{equation}
This estimate can be easily obtained analytically from relation (\ref{QBGI}) if, according to (\ref{costhetab}), we put $\cos\chi = (\Omega R/c)^{1/2}$. It gives
\begin{equation}
P_{\rm cr} \approx 0.15 \, A^{-3/4} \, B_{12}^{1/2}.
\label{Pcr}
\end{equation}

As to the case when the main pulse and interpulse correspond to the same sign of the charge density $\rho_{\rm GJ}$, their death line is to depend on the width of the directivity pattern. In this case, the critical period can be represented with good accuracy as
\begin{eqnarray}
P_{\rm cr}(W_{0} = 3^{\circ}) & \approx & 0.15 \, B_{12}^{1/2}, 
\label{Pcr10} \\
P_{\rm cr}(W_{0} = 5^{\circ}) & \approx & 0.22 \, B_{12}^{1/2}.
\label{Pcr30}
\end{eqnarray}

Thus, we come to important conclusion that the condition of the possibility to observe interpulses 
\begin{equation}
\pi/2 - \chi < \frac{1}{2} \, W_{\rm r} 
\end{equation}
which was usually used is to be corrected. As $\delta W \ll W_{r}$, only a small part of this region corresponds to the possibility to observe the interpulse. As a result, the number of interpulse pulsars turns out to be much less than it has been estimated so far.

\section{Population synthesis}
\label{Sect:popsynt}

\subsection{Kinetic equation method}

Recently we have already analyzed the pulsar distribution on the base of the kinetic equation describing the evolution of neutron stars ~\citep{Arzamassky2017}. As so-called dynamic age of the interpulse pulsars $\tau_{\rm D} = P/2{\dot P}$ is usually small due to their small periods $P < 0.5$ s (see Table~\ref{table0}), it was assumed that the magnetic field for ordinary pulsars can be considered constant. 

Therefore the kinetic equation describing the distribution of pulsars $N(P,\chi, B)$ by period $P$, inclination angle $\chi$ and magnetic field $B$ should be determined from the equation 
\begin{equation}
    \frac{\partial}{\partial P}(\dot{P}N) + \frac{\partial}{\partial \chi}(\dot{\chi} N) = Q,
\label{kinetic}
\end{equation}
where the source $Q$ depends actually on two unknown function, i.e., on initial periods $P$ and inclination angle $\chi$ (as to magnetic field distribution, it can be evaluated from observations). As to the values $\dot{P}$ and $\dot{\chi}$, they should be determined by the specific model of pulsar braking. In particular, for BGI model we have~\citep{BGI93}
\begin{eqnarray}
\dot P & \approx & \frac{\pi f_{\ast}^2}{2} \frac{B_{0}^{2}R^{6}\Omega}{I_{\rm r} c^{3}}\,i_{0}\cos\chi,
\label{BGI_P00}\\ 
\dot\chi & \approx & -\frac{f_{\ast}^2}{4} \frac{B_{0}^{2}R^{6}\Omega^{2}}{I_{\rm r} c^{3}}\, i_{0}\sin\chi.
\label{BGI_chi00}
\end{eqnarray}
Here again $f_{\ast} \sim 1$ is dimensionless polar cap area, $I_{\rm r}$ is the moment of inertia of a star, and $i_{0} = I/I_{\rm GJ}$ is dimensionless electric current circulating in the magnetosphere. These relations can be rewritten as
\begin{eqnarray}
    \dot{P}_{-15} & = & \frac{B_{12}^2}{P}Q_{\rm BGI} \cos^2\chi, 
    \label{BGI_P0}\\
    \dot{\chi} & = &10^{-15} Q_{\rm BGI}\frac{B_{12}^2}{P^2}\sin\chi \cos\chi,
    \label{BGI_chi0}
\end{eqnarray}
where $\dot{P}_{-15} = 10^{15}\dot{P}$, and $Q_{\rm BGI}$ is just the parameter (\ref{QBGI}) entered above. Accordingly, MHD model gives~\citep{TchPhS16}
\begin{eqnarray}
\dot P & \approx & \frac{\pi}{2} \frac{B_{0}^{2}R^{6}\Omega}{I_{\rm r} c^{3}}(1+\sin^{2}\chi),
\label{MHD_P}\\ 
\dot\chi & \approx & -\frac{1}{4} \frac{B_{0}^{2}R^{6}\Omega^{2}}{I_{\rm r} c^{3}} \sin\chi \cos\chi.
\label{MHD_chi}
\end{eqnarray}

Remind that the observable distribution function $N^{\rm obs}(P,\chi, B)$ should be connected with $N(P,\chi, B)$ by relation
\begin{equation}
    N^{\rm obs}(P, \chi, B) = V_{\rm beam}^{\rm vis} V_{\rm lum}^{\rm vis} N(P, \chi, B).
    \label{Nobs}
\end{equation}
Here $V_{\rm lum}^{\rm vis}$ is the luminosity visibility function (we cannot observe far dim objects), and, as before, $V_{\rm beam}^{\rm vis}$ is the beam visibility function depending on the width of the directivity pattern. As to luminosity visibility function $V_{\rm lum}^{\rm vis}$, the standard evaluation is
\begin{equation}
    V_{\rm lum}^{\rm vis} \approx P^{-1}
    \label{VvislumMHD}
\end{equation}
(see~\citealt{TM, Pons, Arzamassky2017} for more detail). Below we use this evaluation for MHD model.

On the other hand, for BGI model we need to correct $V_{\rm lum}^{\rm vis}$ function, as it depends on inclination angle $\chi$, when $\chi \rightarrow 90^{\circ}$. Remember that according to  BGI model~\citep{BGI93} pulsars radio luminosity $L_{\rm rad}$ reach only $10^{-5}$ of the particle energy flux $W_{\rm part}$ near the surface of the neutron star. On the other hand, for fast pulsars ($Q_{\rm BGI} < 1$) we have
\begin{equation}
    W_{\rm part} = Q^2_{\rm BGI}W_{\rm tot}.
\end{equation}
As a result, for space-homogeneous distribution of pulsars the luminosity visibility function $V_{\rm lum}^{\rm vis} \propto L_{\rm rad}^{-1}$ can be presented as
\begin{equation}
    V_{\rm lum}^{\rm vis} = P^{-11/14}B_{12}^{2/7}\cos^{6d-4}\chi.
    \label{Vvislum}
\end{equation}
Finally, for interpulse pulsars the beam visibility function $V_{\rm beam}^{\rm vis}$ (\ref{Vbeam}) is to be changed with the visibility width $\delta W$. 

The convenience of the kinetic approach is that, in the presence of integrals of motion ~\citep{BGI93, TchPhS16}
\begin{eqnarray}
    I_{\rm BGI} & = & \frac{P}{\sin\chi},
\label{IBGI} \\
    I_{\rm MHD} & = & \frac{P\sin\chi}{\cos^2\chi},
    \label{IMHD}
\end{eqnarray}
equation (\ref{kinetic}) can be integrated. Moreover, due to very simple observable distribution $N^{\rm obs}(P) \propto P^{1/2}$ in the domain $0.033 \, {\rm s} < P < 0.5$ s (see~\citealt{Arzamassky2017} for more detail) just overlapping almost all interpulse pulsars this integration can be produced analytically. As a result, it was found that 
\begin{eqnarray}
    N_{\rm MHD}(P, \chi) & = & K_{\rm MHD} \frac{(\pi/2 - \chi - \sin\chi \cos\chi)}{\cos^3\chi}P^2,
    \label{NMHD0}\\
    N_{\rm BGI}(P, \chi) & = & K_{\rm BGI} \frac{(\chi - \sin\chi \cos\chi)}{\sin^3\chi \cos^{2d-1}\chi} P^2,
    \label{NBGI0}
\end{eqnarray}
where again $d \approx 0.75$, and the coefficients $K_{\rm MHD}, K_{\rm BGI}$ are to be determined from the normalization to the entire number of pulsars in the range  $0.03 \, {\rm s} < P < 0.5$ s
\begin{equation}
    N_{\rm tot} = \int_{0.03}^{0.5}{\rm d}P\int_0^{\pi/2}d\chi \, V^{\rm vis}(P, \chi)N(P, \chi).
    \label{Ntot}
\end{equation}
In turn, the number of orthogonal interpulse pulsars in the same range can be determined as
\begin{equation}
    N_{\rm ort} = \int_{0.03}^{0.5}{\rm d}P\int_0^{\pi/2}d\chi 
    \, V_{\rm beam,90}^{\rm vis}(P, \chi)V_{\rm lum}^{\rm vis}(P, \chi)N(P, \chi).
    \label{Nort}
\end{equation}

Note that the question of normalization constant $N_{\rm tot}$, i.e., the total number of isolated pulsars with $0.03 \, {\rm s} < P < 0.5$ s contains a large uncertainty. According to ATNF database~\citep{Manetal2005}, we have $N_{\rm tot} = 968$ (February 2020). But the ATNF catalog is not homogeneous and, in particular, it contains a large number of weak sources, for which the possible interpulse is beyond the sensitivity limit. Thus, hereafter we will consider the \textit{fraction} $f_\mathrm{orth}$ of orthogonal pulsars among overall pulsar population. These fractions can be easy predicted within numerical synthesis. But, on the other hand, they can be evaluated from the statistics of actually observed pulsars. 

The details of estimating of $f_\mathrm{orth}$ are described in Appendix~\ref{AppendixC}. We found that for pulsars with $0.03 \, {\rm s} < P < 0.5$ s there are $f_\mathrm{orth} = 2.5 \div 5.5$\% of orthogonal rotators among the overall population of active pulsars{\footnote{In Table~\ref{table0} we use the normalization for the total number of pulsars in the corresponding range of pulsar periods $P$.}}. Such a wide credible interval is due to relatively small number of actually observed orthogonal pulsars and uncertainties in the procedure to decide if given pulsar is an orthogonal one or not. 

The analysis produced by~\citet{Arzamassky2017} has shown that the number of observable interpulse pulsars with $\chi \approx 0^{\circ}$ can be explained within both BGI (counter-alignment) and MHD (alignment) models due to considerable uncertainties in the initial pulsar distribution $Q(P, \chi)$. In turn, this approach gave us the possibility to evaluate birth functions $Q_{P}(P)$ and $Q_{\chi}(\chi)$ on pulsar initial periods and inclination angles. As was found, for BGI model in the domain $0.03 \, {\rm s} < P < 0.5 \, {\rm s}$ they look like
\begin{eqnarray}
Q_{P}^{\rm BGI}(P) & = & P, \quad
%\label{QPB} \\
Q_{\chi}^{\rm BGI}(\chi) = \frac{2}{\pi}.
\label{QchiB} 
\end{eqnarray}
Accordingly, for MHD model in the domain \mbox{$0.03 \, {\rm s} < P < 0.5 \, {\rm s}$} we have  
\begin{eqnarray}
Q_{P}^{\rm MHD}(P) & = & 1, \quad
%\label{QPM} \\
Q_{\chi}^{\rm MHD}(\chi) =  \sin\chi.
\label{QchiM} 
\end{eqnarray}
As to interpulse pulsars with $\chi \approx 90^{\circ}$, analysis was performed only for MHD model, which also gave the reasonable number of orthogonal interpulse pulsars. 

However, this analysis did not include into consideration the real visibility function for interpulse pulsars \mbox{$V_{\rm beam}^{\rm vis} \approx \delta W$} which, as was shown above, is to diminish drastically the predicted number of orthogonal interpulse pulsars. Nevertheless, below we utilize distribution functions (\ref{NBGI})--(\ref{Iend}) obtained earlier, since the necessary additional corrections to this study refer not so much to the equation itself as to the visibility function $V_{\rm beam}^{\rm vis}$ and features of the behavior of its solution when $\chi \rightarrow 90^{\circ}$. 

Below we assume, as was done by ~\citet{Arzamassky2017}, that pulsar birth function $Q$ can be presented as a product $Q_P(P)Q_\chi(\chi)Q_B(B)$. Indeed, the effects of the distribution over the magnetic field were not taken into account, since integrals of motion (\ref{IMHD})--(\ref{IBGI}) does not depend on magnetic field. This implies that magnetic field can only change the rapidity of the individual pulsar motion along their evolutionary path. Therefore, we can now take into account the distribution on the magnetic field simply by multiplying the previously obtained distribution functions $N(P, \chi)$ by $Q_B(B)/B^{k}$, where according to (\ref{kinetic})--(\ref{MHD_chi}) $k_{\rm BGI} = 10/7$ and $k_{\rm MHD}  =2$. On the other hand, as was shown by ~\citet{BGI93}, with high accuracy one can put
\begin{equation}
    Q_B(B) = \left(\frac{B}{B_n}\right)^a\left(1 + \frac{B}{B_n}\right)^{-1-a-b},
    \label{QB}
\end{equation}
where $B_{0} = 10^{12}\,G$, $a = 2$, and $b = 0.7$. %Разве не b=0.7 ?

As a result, the pulsar distribution function in BGI model $N_{\rm BGI}(P,\chi,B)$ can be written down as
\begin{equation}
    N_{\rm BGI}(P,\chi,B) = K_{\rm BGI} \frac{Q_B(B_{12})}{B_{12}^{10/7}}\frac{(\chi - \sin\chi \cos\chi)}{\sin^3\chi \cos^{2d-1}\chi}P^2,
    \label{NBGI}
\end{equation}
where
\begin{equation}
    K_{\rm BGI} = \frac{N_{\rm tot}}{2W_{0}I_{1}I_{2}I_{3}},
    \label{NBGIB}
\end{equation}
and
\begin{eqnarray}
    I_1 & = & \int_0^{\pi/2}\frac{(\chi - \sin\chi \cos\chi)}{\sin^2\chi}{\rm d}\chi = 1, 
    \label{I11}\\
    I_2 & = & \int_{0.03}^{0.5}P^{5/7}{\rm d}P \approx 0.18, \\
    I_3 & = & \int_0^{\infty}\frac{Q_B(B_{12})}{B_{12}^{8/7}}{\rm d}B_{12} \approx 0.19.
\end{eqnarray}
%In this calculations of the observing distribution function of the interpulse pulsars we should count the geometric visibility function $V_{beam}^{vis}$ in circle angles as also the value of $W_0$
In (\ref{I11}) we put $d = 0.75$.
Accordingly, distribution function in MHD model $N_{\rm MHD}(P,\chi,B)$ looks like
\begin{equation}
    N_{\rm MHD}(P,\chi,B) = K_{\rm MHD} \frac{Q_B(B_{12})}{B_{12}^{2}}
    \frac{(\pi/2 - \chi - \sin\chi \cos\chi)}{\cos^3\chi}P^2,
    \label{NMHD}
\end{equation}
where
\begin{equation}
    K_{\rm MHD} = \frac{N_{\rm tot}}{2W_{0}I_{1}I_{2}I_{3}},
    \label{KMHD}
\end{equation}
and now
\begin{eqnarray}
    I_1 & = & \int_0^{\pi/2} \frac{(\pi/2 - \chi - \sin\chi \cos\chi)\sin\chi}{\cos^3\chi} {\rm d}\chi = \frac\pi4, \\
    I_2 & = & \int_{0.03}^{0.5} P^{1/2} {\rm d}P \approx 0.23, \\
    I_3 & = & \int_0^{\infty}\frac{Q_B(B_{12})}{B_{12}^{2}}{\rm d}B_{12} \approx 0.33.
    \label{Iend}
\end{eqnarray}

\section{Interpulse pulsars as an evolutionary test}
\label{Sect:test}

\subsection{Approximately orthogonal pulsars}
\label{Sect:6.1}

Now we can return to our main goal, i.e., to formulating a test that may clarify the direction of the inclination angle evolution. As was already mentioned above, the central idea is connected with the amount of orthogonal interpulse pulsars, as their number should depend substantially on the sign of the derivative $\dot{\chi}$. For this reason, for orthogonal pulsars the predictions of MHD and BGI model are to differ significantly. 

Remember that according to~\citet{Arzamassky2017} within MHD model the distribution function $N_{\rm MHD}(P, \chi)$ (\ref{NMHD}) reproduces good enough the number of both aligned and orthogonal interpulse pulsars. In particular, the number of orthogonal pulsars in the range $0.03\, {\rm s} < P < \, 0.5 \, {\rm s}$ is $18 \div 40$, in good agreement with observations (see Table 1). However, in this paper, it was supposed that the visibility function of orthogonal interpulse pulsars $V_{\rm beam,90}^{\rm vis}$ is determined by the condition $\pi/2 - \chi \, < \, W_{\rm r}/2$. In Figures~(\ref{fig:7})--(\ref{fig:9}) it corresponds to the complete filling of the area above the dashed line. As was shown in Sect.~\ref{Sect:vis}, this assumption significantly overestimates the number of orthogonal interpulse pulsars.

On the other hand, now the precise accounting for the plasma generation region within magnetic poles, i.e., more accurate determination of the directivity pattern, allows us to specify the number of interpulse pulsars and, hence, to clarify the direction of the inclination angle evolution. In Table~\ref{table3} we present the number of interpulse pulsars $N_{\rm ort, 1}$ (\ref{Nort}) within the domain $0.03 \, {\rm s} < P < 0.5 \, {\rm s} $ for $W_{50} = 3^{\circ}$ and \mbox{$W_{50} = 5^{\circ}$} obtained through the visibility function $V_{\rm beam,90}^{\rm vis} = \delta W$ and distribution functions (\ref{NBGI}) and (\ref{NMHD}). 

We see that for MHD model the number of orthogonal interpulse pulsars is always lower than 1\% Which is barely consistent with $f_\mathrm{orth}$ mentioned above. On the other hand, BGI model predicts larger fraction $1.2 \div 2.3$\% for $A \sim 0.5 \div 0.7$. This probably may indicate that BGI approach is better consistent with the observations. Nevertheless, no choice between two models can be made at this stage.

\begin{table}
    \caption{Comparison of the fraction of approximately orthogonal pulsars $N_{\rm ort,1}$ in percents for different parameters $W_{50}$ and $A$}
    %\vspace{0.3cm}
    \centering
    \begin{tabular}{|c|c|c|c|c|c|c|c|c|}
        \hline 
        $W_{50}$ & \multicolumn{4}{c|}{$3^{\circ}$} & \multicolumn{4}{c|}{$5^{\circ}$} \\ 
        \hline
        $A$            &   0.5  & 0.7 &  1.0  &   1.4  &  0.5 &  0.7  &  1.0 & 1.4    \\
        \hline
        ${\rm MHD}$  &  $0.6$ & $0.4$ & $0.3$ &  $0.1$ & $1.0$ & $0.7$  & $0.5$  &  $0.2$   \\  
%        \hline
        ${\rm BGI}$  &  $1.6$ & $1.2$ & $0.7$ &  $0.4$ & $2.3$ & $1.6$ & $1.1$ &  $0.6$      \\  
        \hline
    \end{tabular}
    \label{table3}
\end{table}

\subsection{BGI correction and exactly orthogonal pulsars}
\label{Sect:6.2}

Here we come to another key subject of our consideration. The point is that the results presented above in Table~\ref{table3} do not allow us to determine the total number of orthogonal interpulse pulsars $N_{\rm ort}$ within BGI model. Indeed, the original version of BGI model determines good enough only one component of the braking torque which is parallel to the magnetic moment. This torque resulting from symmetric (north-south even within polar cap) part of the longitudinal currents $i_{0}$ in (\ref{BGI_P00})--(\ref{BGI_chi00}) vanishes for orthogonal rotator.  As within BGI model individual pulsars evolve to 90 degrees, clarification of the braking law for orthogonal rotator is of particular importance. 

On the other hand, as was shown recently by~\citet{Rashkovetskiy2018}, the braking torque perpendicular to magnetic moment depends on fine details of the distribution of electric currents on the surface of the polar cap, which could not be determined analytically. This has become possible only in recent years, based on the results of numerical modeling~\citep{Spitkovsky06_MHD, kalcont2012, Tchetal2013, Philippov+14_MHD}. As to MHD model, it does not require correction at all, since both the evolution equations (\ref{MHD_P})--(\ref{MHD_chi}) (and, hence, the integral of motion $I_{\rm MHD}$ (\ref{IMHD})) stay true for any inclination angles.

\begin{figure}
    \centering
    \includegraphics[scale=0.35]{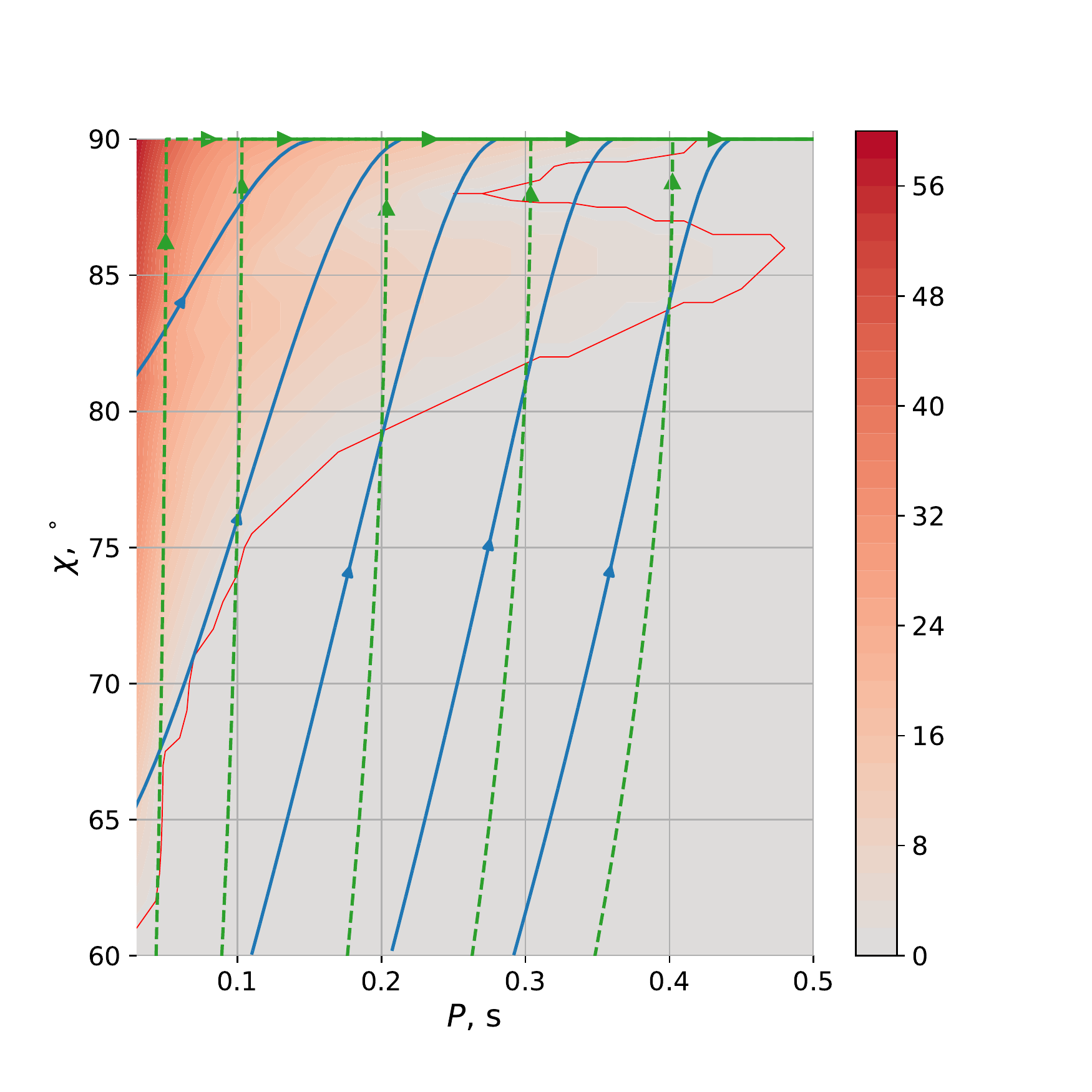}
    \caption{Evolution of individual pulsars for BGI model for $A=0.5$ and $\varepsilon=0.02$. The green dashed trajectories correspond to invariant $I_{\rm BGI}$ (\ref{IBGI}). Blue solid lines are more realistic trajectories defined by (\ref{BGI_P})--(\ref{BGI_chi}) and were used in Monte-Carlo simulation.}
    \label{fig:10}
\end{figure}

As shown in Fig.~\ref{fig:10}, the exact following to invariant $I_{\rm BGI}$ (\ref{IBGI}) leads to unlimited accumulation of pulsars in the region $\chi = 90^{\circ}$. This results from neglecting the second term $C$ in the exact equation of evolution, that in general form can be written down as~\citep{Rashkovetskiy2018} % Beskin2018, %Novoselov2019
\begin{eqnarray}
    \dot{P}_{-15} & = & \frac{B_{12}^2}{P}(Q_{\rm BGI} \cos^2\chi + C), 
    \label{BGI_P}\\
    \dot{\chi} & = &10^{-15} Q_{\rm BGI}\frac{B_{12}^2}{P^2}\sin\chi \cos\chi.
    \label{BGI_chi}
\end{eqnarray}
Here again $B_{12} = B_0/10^{12}$, and $\dot{P}_{-15} = \dot{P}/10^{-15}$. As to extra small factor $C \ll 1$, it just describes the evolution of orthogonal rotators along the line $\chi = 90^{\circ}$.

As was already stressed, the value of $C$ within the framework of the BGI model cannot be determined with the necessary accuracy. In the original work of~\citet{BGI83}, in which only the action of volume magnetospheric currents was taken into account, it was shown that $C$ can be estimated as $(\Omega R/c)\,i_{\rm A}$ where $i_{\rm A} = j_{\parallel}/j_{\rm GJ}$ is the ratio of the longitudinal electric current to the Goldreich-Julian current; recall that in BGI model $i_{\rm A}= 1$. However, as was shown later~\citep{Rashkovetskiy2018}, this estimate did not allow us to explain the pulsar braking  $\dot\Omega$ (\ref{MHD_P}) for $\chi = 90^{\circ}$ within MHD model, because in this model $i_{\rm A} \approx (\Omega R/c)^{-1/2}$ which is too small to get the desired result $C \approx 1$ corresponding to MHD model.

To resolve this contradiction it was assumed that in orthogonal case the energy losses can be connected with additional currents that circulate in magnetosphere along the separatrix separating the areas of open and closed magnetic field lines (see~\citealt{Rashkovetskiy2018} for mode detail). Supposing now that the additional separatrix current is proportional to volume current circulating in the pulsar magnetosphere, i.e. $C = K i_{\rm A}$,  we obtain from the MHD model that $K \approx (\Omega R/c)^{1/2}$. Assuming now that relation $C = K i_{\rm A}$ can be also used for the BGI model, we finally obtain
\begin{equation}
    C \sim \left(\frac{\Omega R}{c}\right)^{1/2}.
    \label{C_correction}
\end{equation}
Since this quantity is not determined with sufficient accuracy, we assume in what follows that
\begin{equation}
    C = \varepsilon \, P^{-1/2}, 
    \label{vare} 
\end{equation}
where the value of $\varepsilon$ belongs to the range between $0.005$ and $0.02$.

As is also shown in Fig.~\ref{fig:10}, for nonzero $C$ the pulsars are to evolve along the axis $\chi = 90^{\circ}$ gradually increasing their period $P$ until they cross the ``death line''. For such pulsars, we can determine the distribution function $ N_ {90} (P, B) $, which satisfies the kinetic equation
\begin{equation}
    \frac{{\rm d}}{{\rm d} P} (\dot{P}_{90} \, N_{90}) 
    = (N_{\rm BGI}{\dot \chi})_{\chi \rightarrow 90^{\circ}}.
    \label{N90_main}
\end{equation}
    Here, according to (\ref{BGI_P}),
\begin{equation}
    \dot{P}_{90} \approx 10^{-15} C \, \frac{B^{2}_{12}}{P}, 
    \label{BGI_P90} 
\end{equation}
and the source in the r.h.s. is determined by the pulsar flow to the region $\chi = 90^{\circ}$ according to (\ref{NBGI}) and (\ref{BGI_chi}).

On the other hand, it is clear that the analytical consideration carried out in Sections~\ref{Sect:6.1}--\ref{Sect:6.2} (corresponding trajectories are shown by green dashed curves in Fig.~\ref{fig:10}) does not allow us to reproduce exactly the evolution trajectory of individual pulsars. Indeed, as shown in Fig.~\ref{fig:10}, sequential consideration of more accurate evolution equations (\ref{BGI_P})--(\ref{BGI_chi}) leads to trajectories presented by blue solid curves which significantly deviate from the trajectories corresponding to conservation of invariant $I_{\rm BGI}$ (\ref{IBGI}) just in the area where orthogonal interpulse pulsars should be observed. More rigorous analysis based on Monte-Carlo simulation is presented in the next Subsection. Here we carry out a qualitative consideration based on the kinetic equation method.

For this reason, as a zero-order estimation we assume that the trajectory reach the boundary $\chi = 90^{\circ}$ not with a period $P$, but with a period $P + P_{0}$, where $P_{0} \approx 0.1$  s (see Fig.~\ref{fig:10}). As a result, solution of the kinetic equation (\ref{N90_main}) 
\begin{equation}
    N_{90}(P,B) =  \frac{1}{{\dot P}_{90}} 
    \int_{0}^{P}(N_{\rm BGI}{\dot \chi})_{\chi \rightarrow 90^{\circ}, P^{\prime} \rightarrow P^{\prime} + P_{0}} \, {\rm d}P^{\prime}  
    \label{BGI_90} 
\end{equation}
on the r.h.s. of which we made a replacement $P \rightarrow P + P_{0}$ looks now like
\begin{equation}
    N_{90}(P,B) = \frac{7 \pi A {\cal K}_{\rm BGI}}{29 \varepsilon} \, 
     \frac{Q_{B}(B_{12})}{B_{12}^2} \, 
    % [(P+P_{0})^{29/14} - P_{0}^{29/14}] \, P^{3/2}.  
     [(P-P_{0})(P + 3P_{0})] \, P^{3/2}.  
    \label{BGI90} 
\end{equation}
Here we neglect the power 1/14. Accordingly, observable distribution function $N_{90}^{\rm obs}(P)$ of such pulsars can be found as
\begin{equation}
    N_{90}^{\rm obs}(P) =  \int_{0}^{\infty} N_{90}(P,B) V_{\rm beam, 90}^{\rm vis} \,
    V_{\rm lum, 90}^{\rm vis} \, {\rm d}B.
    \label{BGI90P} 
\end{equation}
Here the beam visibility function $V_{\rm beam, 90}^{\rm vis}$ is to be equal to $\delta W$, and for luminosity visibility function $V_{\rm lum}^{\rm vis}$ (\ref{Vvislum}) we have to replace $\cos\chi$ with characteristic value $(\Omega R/c)^{1/2}$, so that
\begin{equation}
    V_{\rm lum, 90}^{\rm vis}  \approx 0.1 \, P^{-29/28} B_{12}^{2/7}.
    \label{eq:Vvislum90}
\end{equation}

Finally, it is necessary to stress that according to (\ref{BGI_P90}) magnetic field for orthogonal pulsars $\chi = 90^{\circ}$ is to be estimated as $B_{12} \approx \varepsilon^{-1/2} P^{3/4}{\dot P}_{-15}^{1/2}$, i.e., as
\begin{equation}
    B_{12} \approx 10 \, \varepsilon_{0.01}^{-1/2} \, P^{3/4}{\dot P}_{-15}^{1/2}.
    \label{B90} 
\end{equation}
As to magnetic field distribution function $N_{90}^{\rm obs}(B)$, we obtain
\begin{equation}
    N_{90}^{\rm obs}(B) =  \int_{0}^{\infty} N_{90}(P,B) V_{\rm beam, 90}^{\rm vis} \, V_{\rm lum, 90}^{\rm vis} \, {\rm d}P.
    \label{BGI90B} 
\end{equation}

\begin{table}
    \caption{Comparison of observable (Table~\ref{tableA}) and BGI predicted (\ref{BGI90P}) distributions of orthogonal interpulse pulsars $\chi = 90^{\circ}$ by the period $P$ for $\varepsilon = 0.03$ and $A = 1$}
    \centering
    \begin{tabular}{|c|c|c|c|c|c|}
    \hline
     $P \, [{\rm s}]$  &0.03--0.1 & 0.1--0.2 & 0.2--0.3 &  0.3--0.4 & 0.4--0.5     \\
    \hline
    obs       & $2\div3$ &   $5\div10$   & $7\div8$  &  $3\div4$      & $1\div2$     \\  
     BGI   & 0   &   $6$    & $3$   &   $2$      & $1$ \\ 
    \hline
      \end{tabular}
    \label{table4}
\end{table}

In Table~\ref{table4} we present the comparison of the observable (Table~\ref{tableA}) and predicted (\ref{BGI90P}) distributions of orthogonal interpulse pulsars with $\chi = 90^{\circ}$ by period $P$ for $\varepsilon = 0.03$ and $A = 1$. As we see, the observable distributions is in good agreement with the prediction of BGI model. But, as was already stressed, it was a pretty rough evaluation. Analytical trajectories of the pulsars on $P-\chi$ diagram (see Fig.~\ref{fig:10}), differ from real ones, so there we needed to make a shift of the period, to reduce the artificial amount of orthogonal pulsars with the period less than $0.1$ s. Nevertheless, such a good correlation can be treated as a credible test, that is done for reasonable values of parameters. 

\subsection{Monte-Carlo simulation}
\label{Sect:6.3}

To verify the analytical results presented above, we study the evolution of radio pulsars in the framework of Monte-Carlo approach as well. To reconcile the results of the Monte-Carlo simulations with results obtained within kinetic equation method, we certainly have to use the same birth functions (\ref{QchiB}) and (\ref{QchiM}) of pulsars on periods $Q_{P}(P)$ and inclination angles $Q_{\chi}(\chi)$~\citep{Arzamassky2017}. It is these birth functions that lead to good agreement between the BGI and MHD predictions of the number of aligned interpulse pulsars ${\cal N}_{\rm SP}$ (see Table 1) and observations. Accordingly, the evolution of individual pulsars for BGI model is to be determined by Eqns. (\ref{BGI_P})--(\ref{BGI_chi}), and by Eqns. (\ref{MHD_P})--(\ref{MHD_chi}) for MHD one.

For simplicity, we take into account only three physical parameters of pulsars: period $P$, inclination angle $\chi$ and magnetic field $B$. At the start of simulation a big amount of pulsars is generated according to some initial parameter distribution. After that at each step period $P$ and inclination angle $\chi$ of all existing pulsars are evolved using one of the theoretical models, but magnetic field $B$ is assumed to be constant; also new pulsars are added, the addition rate and parameters are determined by the birth functions from above. We are interested in finding the static distribution so the simulation ran until the number of existing pulsars became close to constant. The initial distribution is not so important for the result as it affects just the convergence speed. However in the BGI case the most convenient option is to start from theoretical $N_{\rm BGI}$ (\ref{NBGI}) with addition of $N_{90}$ (\ref{BGI90}) since the corrections are small. For MHD model the theoretical consideration remains exact since there is no correction, so the corresponding simulations check both the theory and the numerical method.

Then to determine $N_{\rm obs}$ (\ref{Nobs}) within our Monte-Carlo integration method we calculated the sum of all interpulse visibility functions $V_{\rm beam}^{\rm vis}$ (\ref{Vbeam}), $V_{\rm lum}^{\rm vis}$ (\ref{Vvislum}) for BGI model or (\ref{VvislumMHD}) for MHD model to find the proper normalization like in (\ref{NBGIB}) and (\ref{KMHD}). For this, for BGI model we have to take the geometric visibility function $\delta W$ from Fig.~\ref{fig:7}--\ref{fig:9} instead of $V_{\rm beam}^{\rm vis}$ and luminosity visibility function $V_{\rm lum}^{\rm vis}$ (\ref{eq:Vvislum90}) for exactly orthogonal pulsars and (\ref{Vvislum}) for all other ones. For MHD model, where there is no correction, we use the luminosity visibility function (\ref{VvislumMHD}) for orthogonal pulsars as well.

\begin{table}
\caption{Visible orthogonal interpulse pulsars fractions (in percents) with $0.033 \, {\rm s} < P < 0.5 \, {\rm s}$ obtained by Monte-Carlo simulation. Those values that lie within the 
interval $2.5 \div 5.5$\% obtained from observations are
underlined.}
\vspace{0.3cm}
\centering
\begin{tabular}{|c|c|c|c|c|c|c|c|c|}
\hline
$ W_{50} $ & \multicolumn{4}{c|}{$3^\circ$} & \multicolumn{4}{c|}{$5^\circ$} \\
\hline
$A$ & 0.5 & 0.7 & 1 & 1.4 & 0.5 & 0.7 & 1 & 1.4 \\
\hline
MHD & 0.7 & 0.5 & 0.4 & 0.2 & 1.5 & 1.1 & 0.8 & 0.5 \\
\hline
BGI: & & & & & & & & \\
$\varepsilon=0.01$ & 12 & 9.4 & 7.3 & \underline{4.3} & 14 & 11 & 8.3 & \underline{5.0} \\
$\varepsilon=0.02$ & 5.8 & \underline{4.5} & \underline{3.5} & 2.2 & 7.7 & 5.8 & \underline{4.4} & \underline{2.9} \\
$\varepsilon=0.04$ & \underline{3.6} & \underline{2.7} & 1.9 & 1.2 & \underline{5.2} & \underline{3.9} & \underline{2.8} & 1.8 \\
$\varepsilon=1$ & 1.8 & 1.4 & 1.1 & 0.8 & \underline{3.2} & \underline{2.5} & 2.0 & 1.5 \\
\hline
\end{tabular}
\label{table6}
\end{table}

In Table~\ref{table6} we present the fractions of orthogonal interpulse pulsars obtained in Monte-Carlo simulation for different parameters $A$, $W_{0}$, and $\varepsilon$. We see that for reasonable parameters ($A \approx 1$ and $\varepsilon \approx 0.02$) a good agreement with the BGI model can indeed be achieved. On the other hand, we have to stress the strong dependence of the number of orthogonal interpulse pulsars on these parameters. We also noticed that most of visible orthogonal interpulse pulsars in BGI model should have exactly $\chi = 90^\circ$.

\section{Conclusion}

Thus, it was shown that the statistical analysis of orthogonal interpulse pulsars really allows us to formulate a test that can determine the direction of the inclination angle evolution. Two new important points that we included into consideration should be noted. The first one is a significant refinement of the directivity pattern of orthogonal pulsars. In fact, the region of secondary plasma generation near the death line was first determined (cf., e.g.,~\citealt{Qiao, Tsygan}). The second point is the correction to BGI model. We used an updated version of the expression for energy losses for an orthogonal rotator. All the other suggestions (such as the visibility function for ordinary pulsars with inclination angles $\chi \neq 90^{\circ}$, etc.) did not go beyond the standard assumptions commonly used in statistical analysis of radio pulsars. 

As a result, it was shown that  BGI model gives good agreement with observational data; on the other hand, the MHD model predicts too few orthogonal interpulse pulsars. It must be emphasized here that MHD model itself does not require any correction for orthogonal rotators. As for attracting additional opportunities for reconciling the predictions of MHD model with observations (as was done by~\citealt{2015MNRAS.453.3540A} to explain the observed value of the braking index), this work is certainly beyond the scope of this study.

Simply, our result can be explained as follows. In the zero approximation, one can assume that the distribution of the pulsar in the inclination angle $\chi$ weakly depends on this angle. Then using the beam visibility function \mbox{$V_{\rm beam}^{\rm vis} = \sin\chi \, W_{0}$} (\ref{Vbeam}) we can estimate (of course, very roughly) the total number of aligned and orthogonal interpulse pulsars as
\begin{eqnarray}
N_{\rm SP} & \sim &  N_{\rm tot} W_{50}^2, 
\label{NN1} \\
N_{\rm DP} & \sim & N_{\rm tot} W_{50}.
\label{NN2} 
\end{eqnarray}
Here $W_{50}$ is measured in radians. This evaluation indeed gives the reasonable values $N_{\rm DP} \sim 30$--$50$ and $N_{\rm SP} \sim 2$--$5$. But as was stressed in Sect.~\ref{Sect:2}, we can observe orthogonal interpulse pulsars only if their magnetic fields are much larger than those of ordinary radio pulsars. Otherwise, the generation of secondary plasma (and, therefore, the radio emission itself) becomes impossible due to too small potential drop $\psi \propto \rho_{\rm GJ}$ (\ref{psi}) over the pulsar polar cap. As a result, the number of orthogonal interpulse pulsars is to be much smaller than the evaluation (\ref{NN2}) (see Table~\ref{table3}). Only within BGI model which predicts additional class of almost orthogonal pulsars with $\chi \approx 90^{\circ}$, the agreement with observations can be achieved.

In conclusion, it should be noted that in the statistical analysis of the interpulse pulsars we did not take into account the possible correlations associated with the spatial distribution of radio pulsars in the Galaxy, evolution of magnetic field, {\it etc.} which is devoted quite a lot of works (see, e.g.,~\citealt{AChC2002, 2006ApJ...643..332F}).

Finally, it should be noted that the issues discussed above allow us to take a fresh look at many questions arising in the analysis of observations of radio pulsars. For example, the above formula (\ref{B90}) for estimating the magnetic field for almost orthogonal pulsars within BGI model gives much larger values $B_{12} \approx 10 \, P^{3/4}{\dot P}_{-15}^{1/2}$ in comparison with standard estimate $B_{12} \approx P^{1/2}{\dot P}_{-15}^{1/2}$. Accordingly, such pulsars can be observed at much larger periods (of course, unless photon splitting and positronium creation can suppress generation of a secondary plasma, see e.g.~\citealt{MelU, Denis}). In particular, for radio pulsar PSR J0250+5854 ($P = 23.5$ s, ${\dot P}_{-15} = 27$, see~\citealt{23} for more detail) we obtain $B_{12} \approx 300$--$500$, which gives $Q_{\rm BGI} < 1$ even for such enormous pulsar period.

Another example connects with the essential difference of the pair creation domain of orthogonal pulsars compared with standard hollow-cone structure (see Fig.~\ref{fig:caps}). This circumstance must be borne in mind in the analysis  both the mean profiles of radio emission and X-ray radiation.  Indeed, as was already stressed,  up to now one still often assume that for orthogonal pulsars the directivity pattern of radio emission has standard hollow cone structure (see e.g.~\citealt{SiMi}). Moreover, the first results obtained for radio pulsar PSR J0030+0451 by NICER observatory~\citep{NICER1} can be interpreted as if the shape of the heated regions (which  is naturally connected with the region of effective generation of electron-positron plasma) have the crescent shape just as shown in Fig.~\ref{fig:caps}. By the way, according to radio data~\citep{NICER2}, this pulsar is really close to orthogonal, since it has an interpulse exactly between two main pulses.

\section{Acknowledgements}
Authors thank L.I.Arzamasskiy, D.P. Barsukov, A.V.Bilous, A.Jessner, A.A.Philippov and P.Weltevrede for useful discussions. This work was partially supported by Russian Foundation for Basic Research (RFBR), grant 17-02-00788. AB acknowledges the support from the Program of development of M.V. Lomonosov Moscow State University (Leading Scientific School `Physics of stars, relativistic objects and galaxies')

%\bibliographystyle{mnras}
%\bibliography{Novoselov}

\appendix

\section{``Advanced'' list of orthogonal interpulse pulsars}
\label{AppendixA}

\begin{table}
\caption{``Advanced'' list of orthogonal interpulse pulsars
}  
\centering
\begin{tabular}{|c|c|c|c|c|c|c|}
  \hline
 Name        &  $P$    & $\dot P$   & $B_{0}$ & IP/MP  &     Sep.      &   \\
  J          &  [s]    & $10^{-15}$ & $10^{12}$ G & ratio  & [$^{\circ}$]  &          \\  
\hline

0534$+$2200 & 0.033  & 423   &   9.5   &0.6           &145    &    \\
0627$+$0706 & 0.476  & 29.9  &  18.6   &0.2           &180    & $++$   \\
0826$+$2637 & 0.53    &  1.7 &   4.8   &0.01         &180    &    \\
0834$-$4159  & 0.121  & 4.4  &   2.6   &0.25          &171    & $-$     \\
0842$-$4851  & 0.644  & 9.5  &   13.2  &0.14          &180    & $+$     \\
0905$-$5127  & 0.346  & 24.9 &   13.4  &0.06         &175    &  $++$  \\
0908$-$4913  & 0.107  &15.2  &   4.3   &0.24          &176    & $++$    \\
1057$-$5226  & 0.197  & 5.8  &   4.2   &0.5           &205    & $-$     \\
1107$-$5907  & 0.253  &0.09  &   0.6   &0.2           &191    & $-$ \\
1126$-$6054  & 0.203  &0.03  &   0.3   &0.1           &174    & $+$ \\
1244$-$6531  & 1.547  &7.2   &  22.1   &0.3           &145    & $-$  \\
1413$-$6307  &0.395   &7.43  &   8.1   &0.04          &170    & $+$  \\
1549$-$4848  &0.288   &14.1  &   8.8   &0.03          &180    & $++$  \\
1611$-$5209  &0.182   &5.2   &   3.8   &0.1           &177    &     \\
1613$-$5234  &0.655   &6.6   &  11.1   &0.28          &175    &     \\
1627$-$4706  &0.141   &1.7   &   1.8   &0.13          &171    & $-$ \\
1637$-$4553  &0.119   &3.2   &   2.2   &0.1           &173    & $+$ \\
1705$-$1906  &0.299   &4.1   &   4.9   &0.15          &180    & $+$  \\
1713$-$3844  &1.600   &177   &   112   &0.25          &181    &      \\
1722$-$3712  &0.236   &10.9  &   6.7   &0.15          &180    & $++$   \\
1737$-$3555  &0.398   &6.12  &   1.0   &0.04          &180    & $-$  \\
1739$-$2903  &0.323   &7.9   &   7.2   &0.4           &180    & $++$ \\
1828$-$1101  &0.072   &14.8  &   3.1   &0.3           &180    &  $++$  \\
1842$+$0358 &0.233   &0.81   &   1.8   &0.23          &175    &  \\
1843$-$0702  &0.192   &2.1   &   2.5   &0.44          &180    &  \\
1849$+$0409 &0.761   &21.6   &  21.6   &0.5           &181    &    \\
1913$+$0832 &0.134   &4.6    &   2.8   &0.6           &180    &     \\
1915$+$1410 &0.297   &0.05   &   0.5   &0.21          &186    &     \\
1932$+$1059 &0.227   &1.2    &   2.1   &0.02          &170    & $-$  \\
%1935$+$2025 &0.080   &       &         &              &       &   \\
2032$+$4127 &0.143   &20.1   &   6.3   &0.18          &195    & $-$  \\
2047$+$5029 &0.446   &4.2    &   6.6   &0.6           &175    &    \\  
\hline
\end{tabular} 
\label{tableA} 
\end{table}

In table~\ref{tableA} we present ''advanced'' list of orthogonal interpulse pulsars  given by~\citet{Maciesiak2011} including pulsar name, their period $P$, period derivative ${\dot P}$, magnetic field $B_{12}$ evaluated by expression (\ref{B90}) as well as IP/MP radio intensity ratio and IP-MP angular separation. In the last column a plus sign is placed if there is a coincidence with the classification specified by~\citet{MalNik2013}, and a minus in the opposite case\footnote{Note, that classification made by~\citet{MalNik2013} was based on combination of various highly model-dependent methods. Moreover, even calculation errors were found for some pulsars after the paper was published (E. Nikitina, private communication). Therefore, the contradiction mentioned for eight pulsars above should not be considered so earnestly.}. Two pluses are given if there is the confirmation in some other publications.

\section{Determination of magnetic field}
\label{AppendixB}

In this Appendix we remember the procedure of model-consistent estimate of pulsar magnetic field within BGI and MHD approaches. As was shown in Sect.~\ref{Sect3.3}, pulsar death line equation can be rewritten and tested in the form  $\cos^{7/15}\chi > P \cdot ( A^{14/15} B_{12}^{-8/15} )$ (\ref{eq:death_line_angle}). Here $B_{12}$ is model-consistent estimate of pulsar magnetic field taken in \mbox{$10^{12}$ G.}

Within BGI model~\citep{BGI84, BGI93} the determination of magnetic field depends substantially on the parameter $Q_{\rm BGI}$ (\ref{QBGI}). For $Q_{\rm BGI} < 1$ (far from the real ``death line'' where the theory can only be considered as consistent) the spin-down law (\ref{BGI_P0}) results in
\begin{equation}
  B_{12}^{\rm BGI} = A^{-7/10} P^{-1/20} \dot P_{-15}^{7/10} \cos^{-21/20}\chi,
  \label{eq:bgi_field_Q1}
\end{equation}
where $P$ taken in seconds. In particular, for orthogonal rotators the evaluation $B_{12} \approx \varepsilon^{-1/2} P^{3/4}{\dot P}_{-15}^{1/2}$ (\ref{B90}) is to be used. On the other hand, for pulsars in the vicinity of the ``death line'' (i.e., for pulsars with \mbox{$Q_{\rm BGI} > 1$)} the evaluation gives
\begin{equation}
    B_{12}^{\rm BGI} = P^{15/8} \cos^{-1}\chi.
    \label{eq:bgi_field_Q0}
\end{equation}

As to MHD model, corresponding magnetic field is expected to be consistent with the spin-down law (\ref{MHD_P})
\begin{equation}
    P \dot P = \frac{\pi^2 B_{0}^{2} R^{6}}{Ic^3} (1 + \sin^2\chi).
\end{equation}
In other words,
\begin{equation}
    B_{12}^{\rm MHD} = 1.04 \, P^{1/2} P_{-15}^{1/2} (1 + \sin^2\chi)^{-1/2},
\end{equation}
where we assume neutron star radius $R = 12.5$ km and moment of inertia
$I_{\rm r} = 1.5\cdot 10^{45}$ g cm$^2$ \citep{Spitkovsky06_MHD, Philippov+14_MHD}. For orthogonal pulsars within MHD model one can just put $\sin \chi \approx 1$ and get
\begin{equation}
    B_{12}^{\rm MHD,90} = 0.73 \, P^{1/2} P_{-15}^{1/2}.
    \label{B90_MHD}
\end{equation}

\section{ATNF catalogue limitations}
\label{AppendixC}

As it was already stressed, the amount of interpulse pulsars given in Tables~\ref{table0}, \ref{tableA} should be considered as a lower estimate. Indeed, the ATNF catalog is not homogeneous and, in particular, it contains a large number of weak sources, for which the possible interpulse is beyond the sensitivity limit. Below we try to determine the uncertainty
of the normalization constant $N_{\rm tot}$ (\ref{Ntot})
which is important for our analysis.

\begin{figure}
\centering
\includegraphics[scale=0.65]{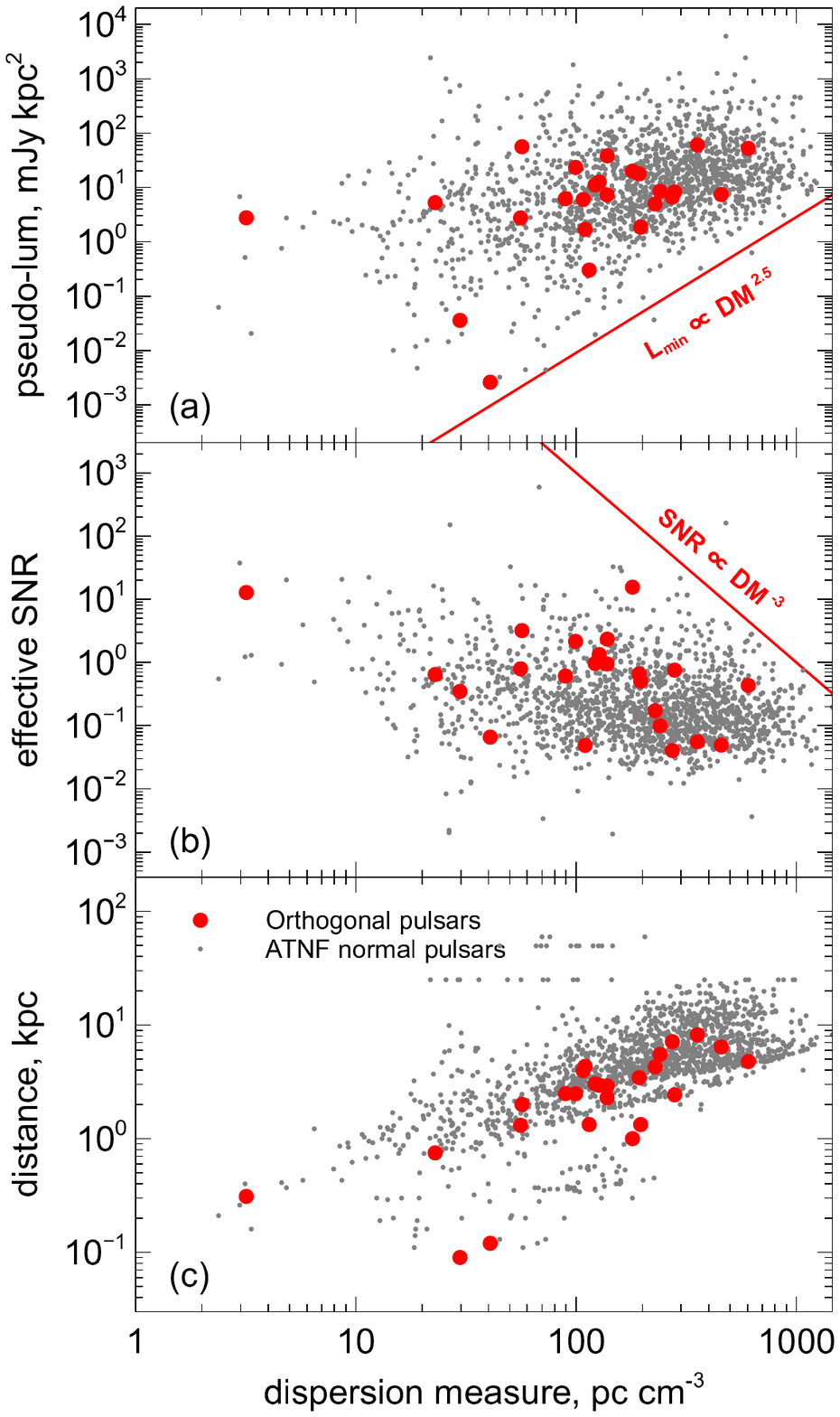}
\caption{
To the discussion of the completeness of the ATNF pulsar database. On the plots (a)-(c) above we show normal radiopulsars included into ATNF which periods are within $0.03\div 0.5$ seconds interval (gray dots). Red circles represent known pulsars with interpulses listed in Table~\ref{tableA}. The description of the plots are as follows: (a) pulsar pseudo-luminosity $L = F_{1.4}\times D^2$, where $F_{1.4}$ is the observed flux at the frequency 1.4 Ghz and $D$ is  distance typically based on the dispersion measure estimation; (b) Effective signal-to-noise ratio (see Eq.\ref{eq:snr}); (c) Dispersion measure-based distance to pulsar. All quantities are plotted against the dispersion measure. We interpret the lack of low-luminosity and high-SNR pulsars with at large $DM \gtrsim 100 $ pc cm$^{-3}$ as a result of the incompleteness of the ANTF database. And assume that this catalogue is more or less complete up to $DM \sim 70-150$ pc cm$^{-3}$.
}
\label{fig:atnf_lim}
\end{figure}

Recall that successful detection of a pulsar within a survey depends on the pulsar radio luminosity, distance and pulse broadening due to interstellar dispersion and distortion. Weak and wide-pulse pulsars are hard to detect at large distances (or at large dispersion measures). This is the reason for the lack of pulsars with low pseudo-luminosities and large dispersion measures on the corresponding diagram, see Figure~\ref{fig:atnf_lim}(a). Grey dots on this plot represent all normal isolated pulsars with periods $0.03\div 0.5$ s stored in the ATNF database, while red circles are for orthogonal pulsars from Table~\ref{tableA}.

One can see that minimal pseudo-luminosity remains approximately constant up to $DM \approx 70$ pc$\cdot$cm$^{-3}$ for ANTF pulsars. While at larger values of $DM$ it scales as $L_{\rm min} \propto DM^{2.5}$. At the same time, there are lack of high-SNR pulsars at  $DM \gtrsim 150$ at Figure~\ref{fig:atnf_lim}(b). The effective signal-to-noise ratio was estimated here as
\begin{equation}
    \mathrm{SNR}_\mathrm{eff} = \frac{F_\mathrm{1.4}}{F_\mathrm{Crab}}\sqrt{
    \frac{P - W_\mathrm{50}}{W_\mathrm{50}}},
    \label{eq:snr}
\end{equation}
where $P$ is pulsar period, $F_{1.4}$ its radio flux at frequency 1.4 GHz, $F_\mathrm{Crab}$ is the Crab pulsar flux at the same frequency and $W_{50}$ is main pulse width at half maximum intensity taken in seconds \citep{johnston17}.

We conclude therefore, that ATNF catalogue can be considered as more or less complete for $DM \lesssim 70\div 150$ pc cm$^{3}$. There are $200\div 400$ isolated normal pulsars in this interval with from 8 to 17
%$3\div 5$ (5-8 minimum) to $10 \div 15$ (12-17 maximum)
orthogonal ones among them correspondingly. Assuming Poissonian statistics for both quantities we finally estimate the fraction of orthogonal pulsars as $f_\mathrm{orth} = \left( 4 \pm 1.5 \right) \%$ 
%$f_\mathrm{orth} = 0.5\div 5$ 
for local galactic volume at $1\sigma$ significance. And since this local pulsar population is merely independent,  then one can expect the same fraction for all radio loud galactic pulsars. In the frame of our work we compare model prediction of $f_\mathrm{orth}$ for both MHD and BGI approaches with the number obtained above.

Note that the above estimate of the fraction of orthogonal interpulse pulsars $f_{\rm orth}$ is in good agreement with another independent estimate which also can be obtained from ATNF catalog.
This criterion connects with discarding pulsars for which ATNF catalogue does not give mean pulse width W10 on the 10\% intensity level. 

Indeed, measured W10 in ATNF catalog indicates that the noise level for a given pulsar is quite low. Hence, one can believe that for such pulsars it is possible to detect an interpulse with a sufficiently large intensity ratio IP/MP. And really, in Table~\ref{tableA}, pulsar PSR J0826+2637 \mbox{(IP/MP = 0.01)} turned out to be the only exception for which ATNF catalog does not give a value of W10. For all other pulsars having IP/MP $>$ 0.03, a value of W10 is provided. 

Hence, if we apply this criterion, the number of observed interpulse pulsars changes only slightly. As for the normalizing total number of pulsars $N_{\rm tot}$ (\ref{Ntot}) in the range \mbox{0.033 s $< P <$ 0.5 s}, their number decreases to 475, i.e. halves in comparison with full ATNF catalogue. According to Table~\ref{table0}, it gives $f_{\rm orth} = \left( 3.6\div 5.2 \right) \%$ in good agreement with the previous estimate.

\end{document}